\documentclass[
twocolumn,
amsmath, amssymb, amsfonts,preprintnumbers,aps,prd,longbibliography
,bibnotes
]{revtex4-2}

\usepackage{graphicx}
\pdfoutput=1

\renewcommand{\sec}[1]{\noindent\textbf{#1. --- }}

\usepackage{dcolumn}
\usepackage{bm}
 \usepackage{amstext}

 \usepackage{color}
 \usepackage{bbold}
 \usepackage{delimset} 
\usepackage[caption=false,justification=justified]{subfig}
\usepackage[hidelinks]{hyperref}
\usepackage{orcidlink}
\usepackage{float}
\usepackage[normalem]{ulem}
\usepackage{mathtools}

\usepackage{tikz}
\usetikzlibrary{decorations.pathmorphing}
\usetikzlibrary{decorations.markings}
\usetikzlibrary{positioning, shapes, snakes}
\usetikzlibrary{arrows.meta}

\usepackage{slashed}

\usepackage{amssymb}

\usepackage[]{shuffle}

\renewcommand{\sec}[1]{\par\noindent\textbf{#1.}\enspace}

\begin{document}

\author{Song He\textsuperscript{a,b}
\!\orcidlink{0000-0002-9534-3844}}
\email{songhe@itp.ac.cn}

\author{Yong Zhang\textsuperscript{c,d}
\!\orcidlink{0000-0002-3522-0885}}
\email{zhangyong1@nbu.edu.cn}

\author{Fan Zhu\textsuperscript{a,e}
\!\orcidlink{0009-0007-2188-4393}}
\email{zhufan25@gscaep.ac.cn}

\author{Pengfei Zhu\textsuperscript{c,d}
\!\orcidlink{0009-0000-2310-947X}}
\email{2511690107@nbu.edu.cn}

\affiliation{
\textsuperscript{a}New Cornerstone Laboratory,
Institute of Theoretical Physics,
Chinese Academy of Sciences,
Beijing 100190, China
\\
\textsuperscript{b}School of Fundamental Physics and Mathematical Sciences,
Hangzhou Institute for Advanced Study and ICTP-AP,
University of Chinese Academy of Sciences,
Hangzhou 310024, China
\\
\textsuperscript{c}Institute of Fundamental Physics and Quantum Technology
\& School of Physical Science and Technology,
Ningbo University, Ningbo, Zhejiang 315211, China
\\
\textsuperscript{d}
Zhejiang Key Laboratory of Extreme Universe, Ningbo University, Ningbo, Zhejiang 315211, China
\\
\textsuperscript{e}Graduate School of China Academy of Engineering Physics,
No.~10 Xibeiwang East Road,
Haidian District,
Beijing 100193, P.R. China
}

\title{String Amplitudes on Integer Rectangles}

\begin{abstract}
Hidden zeros of tree-level string amplitudes of colored scalars lie on
loci specified by integer-valued dimensionless momentum contractions
arranged in rectangles on the kinematic mesh.
We study the same kinematic loci for open-string amplitudes with gluons,
fermions and higher-spin states at arbitrary finite mass levels.
Disk monodromy relations yield \emph{zero-locus factorization}:
at regular points where the monodromy system is invertible, the amplitude
is a weighted sum of physical three-point amplitudes sewn to complete
on-shell $(n-1)$-point amplitudes. Hidden zeros follow when the external
wavefunctions annihilate all contributing three-point couplings.
In the low-energy limit, the massless zero-integer branch recovers
the corresponding field-theory factorization. For closed strings,
including gravitons, persistent cubic zeros in both chiral sectors
remove the disk poles, allowing the Kawai--Lewellen--Tye momentum-kernel
zero to imply amplitude zeros. The operator product expansions of
external vertices give a common worldsheet origin for these conditions
across spins and mass levels. 
\end{abstract}

\maketitle
\sec{Introduction}

\begin{figure}[t]
\centering
\begin{tikzpicture}[font=\small, line cap=round, line join=round]
 \definecolor{rectI}{RGB}{36,102,153}
 \definecolor{rectJ}{RGB}{183,91,29}
 \definecolor{rectPair}{RGB}{105,63,143}
 \begin{scope}[x=0.32cm,y=0.32cm]
  \foreach \r in {1,...,5} {
   \pgfmathtruncatemacro{\smin}{\r+2}
   \foreach \s in {\smin,...,7} {
    \pgfmathsetmacro{\xx}{\s-\r-1}
    \pgfmathsetmacro{\yy}{\r+\s-8}
    \path[fill=black!3,draw=black!28,line width=0.35pt]
     (\xx,{\yy-1}) -- ({\xx+1},\yy) --
     (\xx,{\yy+1}) -- ({\xx-1},\yy) -- cycle;
   }
  }
  \foreach \r in {1,2} {
   \foreach \s in {4,...,7} {
    \pgfmathsetmacro{\xx}{\s-\r-1}
    \pgfmathsetmacro{\yy}{\r+\s-8}
    \path[fill=rectPair!16,draw=rectPair!48,line width=0.4pt]
     (\xx,{\yy-1}) -- ({\xx+1},\yy) --
     (\xx,{\yy+1}) -- ({\xx-1},\yy) -- cycle;
   }
  }
  \draw[rectPair,line width=0.8pt]
   (2,-4) -- (6,0) -- (4,2) -- (0,-2) -- cycle;
  \path[fill=rectPair!42,draw=rectPair,line width=0.75pt]
   (2,-2) -- (3,-1) -- (2,0) -- (1,-1) -- cycle;
  \node[font=\scriptsize] at (2,-1) {$s_{25}$};
  \draw[rectI,line width=1.5pt] (2,-4) -- (0,-2);
  \draw[rectJ,line width=1.5pt] (2,-4) -- (6,0);
  \node[text=rectI] at (0,-3.8) {$I$};
  \node[text=rectJ] at (4.9,-2.5) {$J$};
 \end{scope}
 \node at (0.96,1.98) {kinematic mesh};
 \node at (0.96,-1.98) {$s_{ij}=q_{ij}\in\mathbb Z$};
 \draw[<->,black!60,>=Stealth,line width=0.6pt]
  (2.45,0.15) -- (3.35,0.15);
 \begin{scope}[shift={(5.15,0)}]
  \draw[black!55,line width=0.6pt] (0,0) circle (1.25);
  \draw[rectI,line width=1.5pt]
   (70:1.25) arc[start angle=70,end angle=-20,radius=1.25];
  \draw[rectJ,line width=1.5pt]
   (-70:1.25) arc[start angle=-70,end angle=-245,radius=1.25];
  \draw[rectPair,dashed,line width=0.8pt]
   (0:1.25) -- (-135:1.25);
  \foreach \ang in {45,0}
   \fill[rectI] (\ang:1.25) circle (2.1pt);
  \foreach \ang in {-90,-135,-180,-225}
   \path[fill=rectJ] (\ang:1.25) +(-1.9pt,-1.9pt)
    rectangle +(1.9pt,1.9pt);
  \fill (90:1.25) circle (1.8pt);
  \fill (-45:1.25) circle (1.8pt);
  \node[above] at (90:1.36) {$a=8$};
  \node[above right,text=rectI] at (45:1.30) {$1$};
  \node[right,text=rectI] at (0:1.32) {$i=2$};
  \node[below right] at (-45:1.30) {$b=3$};
  \node[below,text=rectJ] at (-90:1.31) {$4$};
  \node[below left,text=rectJ] at (-135:1.30) {$j=5$};
  \node[left,text=rectJ] at (-180:1.31) {$6$};
  \node[above left,text=rectJ] at (-225:1.30) {$7$};
  \node[text=rectI] at (0.60,0.55) {$I$};
  \node[text=rectJ] at (-0.62,-0.20) {$J$};
 \end{scope}
 \node at (5.15,1.98) {disk ordering};
 \node at (5.15,-1.98) {$({\color{rectI} I},b,{\color{rectJ} J},a)$};
\end{tikzpicture}
\caption{An integer rectangle and its disk ordering, illustrated
for $n=8$, $I=\{1,2\}$, $b=3$, $J=\{4,5,6,7\}$ and $a=8$.
The shaded mesh tiles are precisely $I\times J$; on the disk,
$I$ (circles) and $J$ (squares) occupy boundary arcs separated
by $a,b$. The marked tile and dashed line denote the same
cross pair $(2,5)$, not a propagator in this ordering.}
\label{fig:integer-rectangle-disk}
\end{figure}
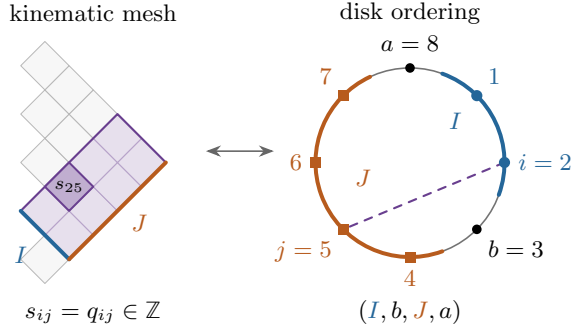

The modern study of hidden zeros began with color-ordered
amplitudes of colored scalars in particle and string theories
\cite{Arkani-Hamed:2023swr}. For $\operatorname{Tr}(\phi^3)$
theory, the nonlinear sigma model, and scalar amplitudes
in Yang--Mills--scalar theory, the zero loci are organized
by rectangles of momentum contractions between two separated
blocks of a color ordering
\cite{Li:2024bwq,Huang:2025blb}.
These pairs form a rectangle in
the kinematic mesh and connect two boundary arcs of the disk
(Fig.~\ref{fig:integer-rectangle-disk}).
The field-theory rectangles have
vanishing entries; their stringy counterparts already include
integer-valued dimensionless invariants.
Factorization near these zeros was also found in
Ref.~\cite{Arkani-Hamed:2023swr}.
Related structures appear
in smooth splitting \cite{Cachazo:2021wsz} and in more
general particle and string splitting relations
\cite{Cao:2024gln,Cao:2024qpp,Arkani-Hamed:2024fyd,
Li:2026cwa}.

This suggests a basic question: what do the same momentum
conditions imply for amplitudes with more general external
states, before imposing any polarization constraints?
We first consider color-ordered amplitudes, allowing gluons,
fermions and higher-spin states throughout the massive string
spectrum \cite{Hanany:2010da}.
Rectangular factorizations in Yang--Mills theory
\cite{Guevara:2024nxd,Zhang:2024efe,ZhangZhu:FermionicNF}
and fermionic hidden zeros \cite{Zhang:2026wye}
provide field-theory instances.
At finite $\alpha'$, the question involves the full physical
string spectrum and its three-point couplings
\cite{Sagnotti:2010at,Schlotterer:2010kk,
Boels:2012if,Markou:2023ffh}:
which fusion channels contribute, and when do their
couplings vanish? 

Disk monodromy relates different color orderings through
momentum-dependent sine factors
\cite{Plahte:1970wy,Bjerrum-Bohr:2009ulz,
Stieberger:2009hq,Boels:2010bv,Ma:2011um}.
Early studies used ordering relations to identify zeros
of dual resonant amplitudes, including excited external
states \cite{DAdda:1971wcy}.
The interplay between monodromy zeros and physical poles
was studied in Ref.~\cite{Boels:2014dka}.
Here we organize the surviving fusion contributions
on general integer rectangles into an $m!\times m!$
sine insertion system, where $m$ is the number of
legs in one block.
Each fusion residue is an on-shell three-point amplitude
sewn to a complete $(n-1)$-point amplitude.
Different cross integers can select different fusion masses;
pairs with no allowed physical intermediate state contribute
no residue.

Under the stated regularity and invertibility conditions,
solving this system gives \emph{zero-locus factorization}:
on the kinematic loci inherited from colored-scalar hidden zeros,
the complete amplitude is a weighted
sum of on-shell three-point amplitudes sewn to complete
$(n-1)$-point amplitudes. The integer condition directly
puts the fused leg at the selected physical mass.
This complements splitting formulas expressed through
off-shell currents \cite{Cao:2024gln,Cao:2024qpp}
and their on-shell representations \cite{Azevedo:2025vxo}.
No polarization bridges are needed for the factorization
formula itself.
Hidden zeros follow when the external polarization,
spinor or tensor wavefunctions annihilate every
contributing physical three-point coupling.
Related integer zeros involving massive bosonic states
and gluinos have been studied using curve integrals
and scaffolding \cite{Chang:2025cqe}.

For closed strings, including gravitons, we use Kawai--Lewellen--Tye (KLT) relations
\cite{Kawai:1985xq,Bjerrum-Bohr:2010pnr}
to combine the rectangular kernel-zero mechanism
\cite{Bartsch:2024amu,Li:2024qfp}
with persistent cubic zeros in both chiral sectors
to obtain amplitude zeros.
The cross operator product expansions (OPEs) give a local
worldsheet perspective on these conditions: their physical
projections encode the selected cubic couplings for bosons
and fermions alike. Under stronger cross-OPE regularity,
the remaining polynomial interactions admit a finite
decomposition into formally scaleless sectors, providing
a relative-scale picture of the zeros.

\medskip
\sec{Zero-locus factorization}
\nopagebreak[4]

To determine the amplitude on the rectangular locus,
keep the external wavefunctions unrestricted and consider
\begin{equation}
 \begin{gathered}
 A_n(I,b,J,a),\qquad I=\{1,\ldots,m\},\\
 b=m+1,\qquad J=(m+2,\ldots,n-1),\qquad a=n,
 \end{gathered}
 \label{eq:sz-blocks}
\end{equation}
with both blocks nonempty. The external states may belong
to arbitrary finite physical mass levels of the open bosonic
string or the flat critical open superstring with the
Gliozzi--Scherk--Olive (GSO) projection. The number of
Ramond external states is even. We use a mostly-plus
metric and define
\begin{equation}
 \begin{gathered}
 k_r^2=-M_r^2,\qquad
 s_{rs}=2\alpha'k_r\cdot k_s,\\
 h_m(\mathbf q):\quad
 s_{ij}=q_{ij}\in\mathbb Z,\qquad i\in I,\ j\in J .
 \end{gathered}
 \label{eq:sz-rectangle}
\end{equation}
The integers may differ between pairs. Momenta obey
conservation and may be complex. Fermionic amplitudes use
a fixed reference ordering of the labelled external fermions,
with statistical signs recorded separately from
Koba--Nielsen phases.

Figure~\ref{fig:integer-rectangle-disk} illustrates
the cross pairs $I\times J$, with $a,b$ separating
the two boundary blocks.
The rectangle constrains only the momenta;
amplitude zeros additionally depend on the
physical fusion couplings.

To approach the rectangle, write
$s_{ij}=q_{ij}+\delta_{ij}$ with fixed integers
$q_{ij}$.
The corresponding fusion data are
\begin{equation}
 \begin{gathered}
 K_{ij}=k_i+k_j,\qquad
 M_{K_{ij}}^2=M_i^2+M_j^2-\frac{q_{ij}}{\alpha'},\\
 \delta_{ij}
 =\alpha'\bigl(K_{ij}^2+M_{K_{ij}}^2\bigr).
 \end{gathered}
 \label{eq:sz-fusion-data}
\end{equation}
Let $\mathcal H_{ij}^{\rm phys}(q_{ij})$ be the complete
allowed physical fusion space at the selected mass and sector.
A pair is spectral when this space is nonempty and
nonspectral otherwise.
For two massless bosonic-string vectors, $q=1$ selects
the tachyon, $q=0$ the massless vector, and $q<0$
massive states; $q\ge2$ is nonspectral.
For massless GSO-projected inputs, positive integers
are nonspectral.
The role of the fusion spectrum in determining integer
roots was emphasized in Ref.~\cite{Boels:2014dka}.

We work in a smooth on-shell neighborhood where the
$\delta_{ij}$ are independent local transverse coordinates,
with holomorphic physical-state representatives.
Apart from the selected cross channels, we exclude
physical poles in all orderings entering the monodromy
relations below. These conditions define the admissible
domain.\footnote{The exclusion includes $P_0^{(i,\rho)}$
and every allowed cyclic consecutive channel, with
complementary subsets identified.
Momenta needed for physical picture isomorphisms
are nonzero \cite{Berkovits:1997mc}.}

Write $J=(j_1,\ldots,j_{|J|})$. For $i\in I$ and
$\rho\in S(I\setminus\{i\})$, introduce
\begin{align}
 P_t^{(i,\rho)}
 &=A_n(\rho,b,j_1,\ldots,j_t,i,j_{t+1},\ldots,j_{|J|},a),
 \nonumber\\
 R_{ij_t}^{\rho}
 &=
 \left.
 \operatorname*{Res}_{\delta_{ij_t}=0}P_t^{(i,\rho)}
 \right|_{h_m(\mathbf q)},
 \label{eq:sz-oriented-residue}
\end{align}
where $0\le t\le |J|$ for $P_t$ and $1\le t\le |J|$
for the residues. The residue is taken before restricting
the remaining cross invariants to the rectangle.
Physical factorization sums over the complete intermediate
physical state space \cite{Boels:2014dka}.
In our sewing conventions, this gives
\begin{equation}
 \begin{aligned}
 &R_{ij_t}^{\rho}
 =\alpha'\tau_{ij_t}\sigma_{ij_t}^{\rho}
 \sum_{\mathsf a,\mathsf b}
 A_3(i,j_t,-K_{ij_t}^{\mathsf b})\,
 (G_{\rm kin}^{-1})_{\mathsf b\mathsf a}\\
 &\quad\times
 A_{n-1}(\rho,b,j_1,\ldots,j_{t-1},
 K_{ij_t}^{\mathsf a},j_{t+1},\ldots,j_{|J|},a)
 \big|_{h_m(\mathbf q)} .
 \end{aligned}
 \label{eq:sz-physical-sewing}
\end{equation}
The labels $\mathsf a,\mathsf b$ index bases of the
complete selected physical state spaces at momenta
$K_{ij_t}$ and $-K_{ij_t}$, respectively.
Their nondegenerate kinetic pairing $G_{\rm kin}$
enters the channel propagator
$G_{\rm kin}^{-1}/(K_{ij_t}^2+M_{K_{ij_t}}^2)
=\alpha'G_{\rm kin}^{-1}/\delta_{ij_t}$;
see Refs.~\cite{Kunitomo:2015usa,Kunitomo:2016bhc}
for Neveu--Schwarz (NS) and Ramond prescriptions.
Our orientation factor is
\begin{equation}
 \tau_{ij}=-(-1)^{q_{ij}+\mathfrak p_i\mathfrak p_j},
 \label{eq:sz-sewing-orientation}
\end{equation}
where $\mathfrak p_i=0$ for bosonic or NS states
and $\mathfrak p_i=1$ for Ramond states.
The sign $\sigma_{ij_t}^{\rho}$
converts the parent and children to their fermion
reference orders; it equals one for bosonic states.\footnote{Explicitly,
$\sigma_{ij_t}^{\rho}=\chi_F(P_t)\chi_3\chi_{n-1}$,
where each $\chi$ is the permutation sign from a fixed
fermion reference order to the displayed word.
Each child inherits the surviving external reference
subsequence and appends its internal fermion label last,
when present.}
Both sewn factors are on-shell amplitudes, with
the fused leg at the selected physical mass.
The $(n-1)$-point factor is complete.
Nonspectral pairs have zero physical residue.

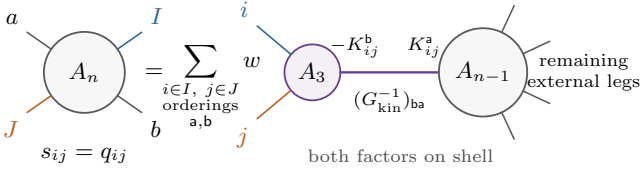
\begin{figure}[t]
\centering
\resizebox{\columnwidth}{!}{%
\begin{tikzpicture}[
 font=\small,
 line cap=round,
 line join=round,
 amplitude/.style={
  draw=black!65,
  fill=black!3,
  line width=0.7pt
 },
 external/.style={
  draw=black!65,
  line width=0.6pt
 }
]
 \definecolor{sewI}{RGB}{36,102,153}
 \definecolor{sewJ}{RGB}{183,91,29}
 \definecolor{sewFusion}{RGB}{105,63,143}

 \begin{scope}[shift={(0.75,0)}]
  \draw[amplitude] (0,0) circle (0.55);
  \node at (0,0) {$A_n$};

  \draw[external,sewI]
   (35:0.55) -- (35:0.95)
   node[above right,text=sewI] {$I$};
  \draw[external]
   (-35:0.55) -- (-35:0.95)
   node[below right] {$b$};
  \draw[external,sewJ]
   (-145:0.55) -- (-145:0.95)
   node[below left,text=sewJ] {$J$};
  \draw[external]
   (145:0.55) -- (145:0.95)
   node[above left] {$a$};

  \node at (0,-1.13) {$s_{ij}=q_{ij}$};
 \end{scope}

 \node at (1.7,0) {$=$};

 \node at (2.48,-.2) {%
  $\displaystyle
   \sum_{\substack{
    i\in I,\ j\in J\\
    \text{orderings}\\
    \mathsf a,\mathsf b
   }} w$
 };

 \begin{scope}[shift={(3.85,0)}]
  \draw[
   amplitude,
   draw=sewFusion,
   fill=sewFusion!8
  ] (0,0) circle (0.38);
  \node at (0,0) {$A_3$};

  \draw[external,sewI]
   (140:0.38) -- (140:0.98)
   node[above left,text=sewI] {$i$};
  \draw[external,sewJ]
   (-140:0.38) -- (-140:0.98)
   node[below left,text=sewJ] {$j$};
 \end{scope}

 \draw[sewFusion,line width=0.9pt]
  (4.23,0) -- (5.57,0);
 \node[above,font=\scriptsize]
  at (4.43,0.08) {$-K_{ij}^{\mathsf b}$};
 \node[above,font=\scriptsize]
  at (5.37,0.08) {$K_{ij}^{\mathsf a}$};
 \node[below,font=\scriptsize]
  at (4.90,-0.10)
  {$(G_{\rm kin}^{-1})_{\mathsf b\mathsf a}$};

 \begin{scope}[shift={(6.17,0)}]
  \draw[amplitude] (0,0) circle (0.6);
  \node at (0,0) {$A_{n-1}$};

  \foreach \ang in {65,25,-25,-65} {
   \draw[external]
    (\ang:0.6) -- (\ang:1.0);
  }
  \node[align=center,font=\scriptsize]
   at (1.37,0) {remaining\\external legs};
 \end{scope}

 \node[font=\scriptsize,text=black!70]
  at (5.05,-1.13) {both factors on shell};
\end{tikzpicture}%
}
\caption{Zero-locus factorization:
the complete restricted amplitude is a weighted sum of
on-shell three-point couplings sewn to complete
$(n-1)$-point amplitudes. The weights $w$ contain the
inverse sine insertion matrix, transport signs and
normalization factors. The internal line denotes the
physical-state pairing, with no propagator denominator.}
\label{fig:zero-locus-sewing}
\end{figure}

Define
$(\boldsymbol A_{\mathbf q})_\omega
=A_n(\omega,b,J,a)|_{h_m(\mathbf q)}$
for $\omega\in S(I)$.
On the admissible domain, these amplitudes satisfy
\begin{equation}
 \boxed{
 \begin{gathered}
 \mathsf S_{\sin}\boldsymbol A_{\mathbf q}
 =\pi\boldsymbol T_{\mathbf q},\\
 (\boldsymbol T_{\mathbf q})_{(i,\rho)}
 =\sum_{t=1}^{|J|}(-1)^{c_{i,t}}R_{ij_t}^{\rho},
 \qquad
 c_{i,t}=\sum_{u=1}^{t}q_{ij_u}.
 \end{gathered}
 }
 \label{eq:sz-residue-system}
\end{equation}
The sine insertion matrix has entries
\begin{equation}
 (\mathsf S_{\sin})_{(i,\rho),\omega}
 =
 \begin{cases}
 \displaystyle
 \sin\pi\left(s_{ib}+\sum_{h>_\omega i}s_{ih}\right),
 & \omega\setminus i=\rho,\\
 0, & \text{otherwise},
 \end{cases}
 \label{eq:sz-insertion-matrix}
\end{equation}
where $\omega\setminus i$ deletes $i$ from $\omega$,
and $h>_\omega i$ means that $h$ lies to its right.
Rows are labelled by $(i,\rho)$ and columns by
$\omega\in S(I)$, giving an $m!\times m!$ matrix.
Where its determinant is nonzero on the physical
kinematic family, zero-locus factorization reads
\begin{equation}
 \boxed{
 \boldsymbol A_{\mathbf q}
 =\pi\mathsf S_{\sin}^{-1}\boldsymbol T_{\mathbf q}.
 }
 \label{eq:sz-residue-factorization}
\end{equation}
The original ordering is the component
$\omega=(1,\ldots,m)$.
Together with Eq.~\eqref{eq:sz-physical-sewing}, this gives
the on-shell $3\times(n-1)$ reconstruction illustrated
in Fig.~\ref{fig:zero-locus-sewing}.
It is an identity for the finite restricted amplitude,
not the residue of a pole in the original ordering.
The matrix has the same form across finite physical
string levels; its sources contain the selected
physical cubic couplings sewn to complete lower-point
amplitudes. Nonspectral pairs have zero residue,
but their integers still enter the cumulative signs
$(-1)^{c_{i,t}}$ multiplying later spectral
contributions.

To derive Eq.~\eqref{eq:sz-residue-system}, use
complete physical disk monodromy
\cite{Bjerrum-Bohr:2009ulz,Stieberger:2009hq,
Boels:2010bv,Ma:2011um}.
Let $\omega_r$ insert $i$ after the first $r$
entries of $\rho$, and set
\begin{equation}
 Q_{i,r}=s_{ib}+\sum_{u>r}s_{i\rho_u},
 \qquad
 C_{i,t}=\sum_{u=1}^{t}s_{ij_u}.
 \label{eq:sz-insertion-charges}
\end{equation}
Before restricting to the rectangle,
\begin{equation}
 \sum_{r=0}^{m-1}\sin(\pi Q_{i,r})
 A_n(\omega_r,b,J,a)
 =\sum_{t=1}^{|J|}\sin(\pi C_{i,t})P_t^{(i,\rho)}.
 \label{eq:sz-full-monodromy}
\end{equation}
The mass-shell shifts $2\alpha'M_i^2\in2\mathbb Z$
leave these sine coefficients unchanged.

The joint integer limit combines neighboring
pole contributions. Physical factorization gives
\begin{equation}
 \operatorname*{Res}_{\delta_{ij_v}=0}P_{v-1}
 =-(-1)^{q_{ij_v}}
 \operatorname*{Res}_{\delta_{ij_v}=0}P_v,
 \qquad v\ge2,
 \label{eq:sz-residue-reversal}
\end{equation}
as a meromorphic identity along the entire divisor.
Suppressing row labels, define the rephased tail
$\widetilde B_t=\sum_{u=t}^{|J|}(-1)^{c_{i,u}}P_u$.
Selected cross-pair poles occur only at the moving
leg's neighbors, so Eq.~\eqref{eq:sz-residue-reversal}
cancels all its poles except the possible simple pole
at $\delta_{ij_t}=0$.
Its regular remainder is jointly holomorphic.

Summation by parts pairs each tail with
$\sin(\pi d_{i,t})-\sin(\pi d_{i,t-1})$,
where $d_{i,t}=\sum_{u=1}^{t}\delta_{ij_u}$
and $d_{i,0}=0$.
This difference is divisible by $\delta_{ij_t}$,
with jointly holomorphic quotient
\begin{equation}
 \frac{\sin(\pi d_{i,t})-\sin(\pi d_{i,t-1})}
 {\delta_{ij_t}}
 \longrightarrow\pi.
 \label{eq:sz-sine-quotient}
\end{equation}
Each surviving pole contributes
$\pi(-1)^{c_{i,t}}R_{ij_t}^{\rho}$, while the
regular remainders vanish in the limit.
Thus the right-hand side of
Eq.~\eqref{eq:sz-full-monodromy} tends jointly to
$\pi\sum_{t=1}^{|J|}(-1)^{c_{i,t}}R_{ij_t}^{\rho}$,
independently of the relative approach rates.
The left-hand side becomes
$\mathsf S_{\sin}\boldsymbol A_{\mathbf q}$,
proving Eq.~\eqref{eq:sz-residue-system}.

Details of the rectangular limit and explicit
four- and six-point factorization formulas, including
examples with massive bosonic and Ramond external
states, are given in the Supplemental Material (SM).

\medskip
\sec{Fusion zeros and the field-theory limit}

The integers select the physical fusion channels,
while external wavefunctions determine their couplings.
On the admissible domain with
$\det\mathsf S_{\sin}\ne0$, suppose that every spectral
cross pair satisfies
\begin{equation}
 A_3(i,j,-K_{ij}^{\eta})=0
 \qquad
 \forall\eta\in\mathcal H_{ij}^{\rm phys}(q_{ij}).
 \label{eq:hz-all-fusion-couplings}
\end{equation}
Complete physical sewing then gives
\begin{equation}
 \boxed{
 \boldsymbol T_{\mathbf q}=0
 \quad\Longrightarrow\quad
 \boldsymbol A_{\mathbf q}=0 .
 }
 \label{eq:hz-open-fusion-zero}
\end{equation}
These conditions need hold only at the rectangular
point, since the lower amplitudes and inverse physical
pairing are regular there.
Fully nonspectral rectangles require no additional
state conditions.
The criterion is sufficient: cancellations among
nonzero fusion contributions can also produce zeros.

For two massless gluons at $q_{ij}=0$, the selected
massless-vector cubic functional vanishes under
\begin{equation}
 \epsilon_i\cdot k_j
 =k_i\cdot\epsilon_j
 =\epsilon_i\cdot\epsilon_j=0,
 \qquad i\in I,\ j\in J .
 \label{eq:hz-massless-gg}
\end{equation}
These bridges apply in both the bosonic string and
the superstring, including the finite-$\alpha'$
bosonic contribution
\cite{Sagnotti:2010at,Schlotterer:2010kk}.
At $q_{ij}=1$, bosonic gluons instead select the
tachyon, whose cubic coupling vanishes when
\begin{equation}
 \epsilon_i\cdot\epsilon_j
 -2\alpha'
 (\epsilon_i\cdot k_j)(k_i\cdot\epsilon_j)=0.
 \label{eq:hz-gluon-tachyon}
\end{equation}
Equation~\eqref{eq:hz-massless-gg} therefore also
removes this channel.

For two first-massive bosonic tensors,
$M_i^2=M_j^2=1/\alpha'$, let $E_i,E_j$ be their
symmetric, transverse and traceless polarizations.
At $q_{ij}=2,3$, they select the massless vector
and tachyon, respectively.
The standard tensor vertices \cite{Sagnotti:2010at}
give the sufficient conditions
\begin{equation}
 E_i\cdot k_j=E_j\cdot k_i=0,
 \qquad E_i:E_j=0.
 \label{eq:hz-first-massive-tensors}
\end{equation}
Here $E_i:E_j:=E_{i,\mu\nu}E_j^{\mu\nu}$, and the
momentum contractions vanish in every tensor slot.
Nonzero partial contractions between the tensors
are allowed.

In the ten-dimensional open superstring, a massless
Ramond state $f_i$ and a first-massive Ramond state
$F_j$, with $M_j^2=1/\alpha'$, select a massless NS
vector at $q_{ij}=1$.
Let $v_j^\mu$ be the independent chiral, transverse
and gamma-traceless vector-spinor polarization of $F_j$.
The complete cubic coupling \cite{Schlotterer:2010kk}
is proportional to
$\epsilon^\eta_{K_{ij},\mu}\mathcal J_{ij}^{\mu}$, where
\begin{equation}
 \mathcal J_{ij}^{\mu}
 =
 \bar\chi_i\slashed k_jv_j^\mu
 -\bar\chi_i\Gamma^\mu(k_{i,\nu}v_j^\nu),
 \label{eq:hz-first-massive-ramond-current}
\end{equation}
with $\bar\chi_i=\chi_i^TC$ and
$\slashed k_j=k_{j,\mu}\Gamma^\mu$.
Setting $\mathcal J_{ij}^{\mu}=0$ suffices;
a stronger choice is
\begin{equation}
 k_{i,\mu}v_j^\mu=0,\qquad
 \bar\chi_i\slashed k_jv_j^\mu=0
 \quad\text{for every }\mu.
 \label{eq:hz-first-massive-ramond}
\end{equation}
These cubic conditions require no complete cross-OPE
regularity.
Derivations, leading-Regge extensions and massless
spinor bridges \cite{Zhang:2026wye,ZhangZhu:FermionicNF} are given in the SM.

For the massless zero-integer branch, take
$\alpha'\to0$ at fixed external momenta.\footnote{Fixed
nonzero cross integers and fixed massive string levels
lie outside this fixed-momentum low-energy limit.}
Define $\widehat s_{ij}:=k_i\cdot k_j$, so that
$s_{ij}=2\alpha'\widehat s_{ij}$.
The zero-integer rectangle already imposes
$\widehat s_{ij}=0$ for every cross pair.
Expanding the sine entries gives
\begin{equation}
 \begin{aligned}
 \mathsf S_{\sin}
 &=2\pi\alpha'\mathsf S_{\rm lin}+O(\alpha'^3),\\
 (\mathsf S_{\rm lin})_{(i,\rho),\omega}
 &=
 \begin{cases}
 \displaystyle
 \widehat s_{ib}
 +\sum_{h>_\omega i}\widehat s_{ih},
 & \omega\setminus i=\rho,\\
 0, & \text{otherwise}.
 \end{cases}
 \end{aligned}
 \label{eq:sz-ft-linear-matrix}
\end{equation}
For the selected massless channels,
$\delta_{ij}=\alpha'K_{ij}^2$ and
$R_{ij}^{\rho}=\alpha'r_{ij}^{\rho}$,
where $r_{ij}^{\rho}$ is the physical $K_{ij}^2$
residue taken before restricting the remaining cross
invariants, as in Eq.~\eqref{eq:sz-oriented-residue}.
All cumulative transport signs equal one at $q_{ij}=0$.
Dividing Eq.~\eqref{eq:sz-residue-system} by
$2\pi\alpha'$ and taking the regular low-energy
limit after separating these poles yields
\begin{equation}
 \mathsf S_{\rm lin}\boldsymbol A^{\rm FT}
 =\frac12\boldsymbol r^{\rm FT},
 \qquad
 (\boldsymbol r^{\rm FT})_{(i,\rho)}
 =\sum_{j\in J}r_{ij}^{\rho,{\rm FT}}.
 \label{eq:sz-ft-system}
\end{equation}
The superscript $\mathrm{FT}$ denotes field-theory
limits in the corresponding Yang--Mills normalization;
the factor $1/2$ follows from $K_{ij}^2=2\widehat s_{ij}$.
Where $\mathsf S_{\rm lin}$ is invertible, this recovers
the on-shell rectangular factorization of Yang--Mills
amplitudes \cite{Guevara:2024nxd,Zhang:2024efe}
and its super-Yang--Mills counterpart
\cite{ZhangZhu:FermionicNF}.

\medskip
\sec{Closed-string extension}

The open-string cubic-zero conditions also give
zeros for closed-string amplitudes, including
graviton amplitudes, when they hold persistently
in both chiral sectors.\footnote{For sums of factorized
chiral states, the sufficient conditions are imposed
term by term.}
We use the closed-string variables
$s^c_{ij}=(\alpha'_c/2)k_i\cdot k_j$,
with $\delta_{ij}=s^c_{ij}-q_{ij}$.
For level-matched physical states of the closed
bosonic or flat critical Type-II string, impose
\begin{equation}
 \begin{gathered}
 \left.A_{3,X}(i,j,-K_{ij}^{\eta})
 \right|_{\delta_{ij}=0}\equiv0,
 \qquad X=L,R,\\
 \forall\eta\in
 \mathcal H_{ij,X}^{\rm phys}(q_{ij}),
 \end{gathered}
 \label{eq:klt-two-chiral-cubics}
\end{equation}
for every spectral cross pair. Then, under the
regularity assumptions below,
\begin{equation}
 \boxed{
 \left.\mathcal M_n\right|_{\rm integer\ rectangle}=0 .
 }
 \label{eq:klt-sphere-zero}
\end{equation}

The proof uses KLT
\cite{Kawai:1985xq,Bjerrum-Bohr:2010pnr},
\begin{equation}
 \mathcal M_n\propto
 \boldsymbol A_L^T\mathcal K\boldsymbol A_R,
 \label{eq:klt-bilinear}
\end{equation}
where the complete chiral disk vectors use
$\alpha'_o=\alpha'_c/4$.
The field-theory rectangular kernel-zero argument
\cite{Bartsch:2024amu,Li:2024qfp}
extends to the sine kernel on integer rectangles:
in appropriately adapted complete bases,
\begin{equation}
 \left.\mathcal K\right|_{\rm integer\ rectangle}=0
 \label{eq:klt-rectangle-kernel-zero}
\end{equation}
entrywise. The bases and the short kernel proof
are given in the SM.

Either disk vector can have fusion poles that
compensate the kernel zero
\cite{Boels:2014dka}.
Equation~\eqref{eq:klt-two-chiral-cubics} removes
these poles by physical factorization.
The cubic identities must hold along each individual
fusion divisor, with the other parameters free;
vanishing only at the rectangular intersection
does not suffice.
On a smooth physical family with independent
$\delta_{ij}$, holomorphic state representatives
and nondegenerate physical pairings, exclude
all additional physical poles in both complete
KLT word sets.\footnote{The exclusion covers
every allowed cyclic consecutive channel,
with complementary subsets identified.
Momenta needed for physical picture isomorphisms
are nonzero.}
Both disk vectors are then jointly regular,
so the kernel zero annihilates the remaining
bilinear. Their regular values need not vanish.
The argument does not require the open-string
factorization formula or invertibility
of $\mathsf S_{\sin}$.
The fusion-pole removal mechanism is illustrated
at four points in the SM.

\medskip
\sec{Cross OPEs and relative scaling}

The cross OPEs provide a local worldsheet perspective
on the fusion-zero conditions: their on-shell physical
projections encode the selected three-point couplings.
Rectangular kinematics can considerably simplify
their singular structure, as illustrated by the
wavefunction bridges governing massless fermionic
fusion on the zero-integer branch
\cite{Zhang:2026wye,ZhangZhu:FermionicNF}.
A relative-scale interpretation becomes available
under the stronger condition that the complete cross
OPEs have no singular part:
\begin{equation}
 \operatorname{PP}_{z_i=z_j}(V_iV_j)=0,
 \qquad i\in I,\quad j\in J,
 \label{eq:ws-cross-regularity}
\end{equation}
where $\operatorname{PP}$ denotes the principal part.
The full OPE includes the Koba--Nielsen (KN) factor
$(z_i-z_j)^{q_{ij}}$ generated by the exponential
vertex factors, together with the chosen ghost
and picture contributions.
For closed strings, impose this condition separately
in both chiral sectors.\footnote{When picture-changing
operators are required, their contributions are included
in both the cross-regularity condition and the
relative-scale analysis.}

\begin{figure}[t]
\centering
\resizebox{\columnwidth}{!}{
\begin{tikzpicture}[
 font=\small,
 line cap=round,
 line join=round,
 leftpt/.style={circle,fill=dilateI,inner sep=1.8pt},
 rightpt/.style={rectangle,fill=dilateJ,inner sep=1.8pt},
 anchorpt/.style={circle,fill=black,inner sep=2pt},
 scaleaxis/.style={black!65,-{Latex[length=1.8mm]},line width=0.6pt},
 dilate/.style={dilateJ,-{Latex[length=1.8mm]},line width=0.7pt}
]
 \definecolor{dilateI}{RGB}{36,102,153}
 \definecolor{dilateJ}{RGB}{183,91,29}

 \foreach \row/\scale/\tag in {0/1.8/1,1.5/3.0/2} {
  \draw[scaleaxis] (-3.25,\row) -- (4.15,\row);
  \node[anchor=east] at (-3.40,\row) {$\lambda_{\tag}$};

  \node[leftpt] at (-2.80,\row) {};
  \node[leftpt] at (-1.90,\row) {};
  \node[leftpt] at (-1.00,\row) {};
  \node[anchorpt] at (0,\row) {};

  \foreach \shape in {0.30,0.65,1.00} {
   \node[rightpt] at ({\scale*\shape},\row) {};
  }
 }

 \node[below,text=dilateI] at (-2.80,-0.10) {$1$};
 \node[below,text=dilateI] at (-1.90,-0.10) {$\cdots$};
 \node[below,text=dilateI] at (-1.00,-0.10) {$m$};
 \node[below] at (0,-0.10) {$b$};
 \node[below,text=dilateJ] at (0.54,-0.10) {$j_1$};
 \node[below,text=dilateJ] at (1.17,-0.10) {$\cdots$};
 \node[below,text=dilateJ] at (1.80,-0.10) {$j_{|J|}$};
 \node[below] at (3.85,-0.10) {$z_a=\infty$};

 \node[above,text=dilateI] at (-2.80,1.60) {$1$};
 \node[above,text=dilateI] at (-1.90,1.60) {$\cdots$};
 \node[above,text=dilateI] at (-1.00,1.60) {$m$};
 \node[above] at (0,1.60) {$b$};
 \node[above,text=dilateJ] at (0.90,1.60) {$j_1$};
 \node[above,text=dilateJ] at (1.95,1.60) {$\cdots$};
 \node[above,text=dilateJ] at (3.00,1.60) {$j_{|J|}$};
 \node[above] at (3.85,1.60) {$z_a=\infty$};

 \draw[dilate] (0.54,0.18) -- (0.90,1.32);
 \draw[dilate,dashed,opacity=0.65] (1.17,0.18) -- (1.95,1.32);
 \draw[dilate] (1.80,0.18) -- (3.00,1.32);
 \node[text=dilateJ,align=center] at (3.60,0.75)
  {$\lambda_1x_t\mapsto\lambda_2x_t$};
 \node[text=dilateI] at (-1.90,0.75) {$I\ \text{fixed}$};
 \node[font=\scriptsize] at (-1.15,-0.63) {$z_m=-1$};
 \node[font=\scriptsize] at (0.08,-0.63) {$z_b=0$};
 \node[text=dilateJ] at (1.17,-0.63) {$J$};

 \node[align=center,font=\scriptsize] at (0.25,-1.23)
  {$s_{ij}=q_{ij}\in\mathbb Z,\qquad
    \operatorname{PP}_{z_i=z_j}(V_iV_j)=0$\\[2pt]
   $(i\in I,\ j\in J)$};
\end{tikzpicture}%
}
\caption{Relative dilation of the $J$ punctures with
$z_b=0$, $z_a=\infty$ and the $I$ punctures fixed.
Writing $z_{j_t}=\lambda x_t$ separates the relative
scale $\lambda$ from the block shapes.
On an integer rectangle, complete cross-OPE regularity
leaves polynomial cross interactions, which decompose
into finitely many homogeneous scale sectors.}
\label{fig:hidden-zero-dilation}
\end{figure}
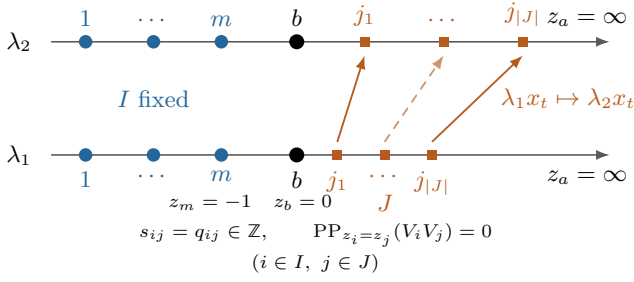

To isolate the relative scale, write
$J=(j_1,\ldots,j_{|J|})$ and fix
$z_b=0$, $z_a=\infty$, $z_m=-1$.
Parameterize the ordered chamber by
\begin{equation}
 \begin{gathered}
 z_i=-y_i,\qquad y_1>\cdots>y_m=1,\\
 z_{j_t}=\lambda x_t,\qquad
 0<x_1<\cdots<x_{|J|}=1,\qquad\lambda>0.
 \end{gathered}
 \label{eq:ws-scale-coordinates}
\end{equation}
As illustrated in Fig.~\ref{fig:hidden-zero-dilation},
varying $\lambda$ dilates the entire $J$ block about
$z_b=0$, while the $I$ punctures remain fixed.
The variables $x,y$ describe the block shapes,
whose domain is independent of $\lambda$.

The finite-puncture KN factors
\cite{Koba:1969rw} become
\begin{equation}
 \begin{gathered}
 \mathrm{KN}_{\rm rect}
 =\lambda^{\beta_J}
 \mathrm{KN}_I(y)\,\mathrm{KN}_J(x)
 \prod_{i\in I}\prod_{t=1}^{|J|}
 (\lambda x_t+y_i)^{q_{ij_t}},\\
 \beta_J=\sum_{t=1}^{|J|}s_{bj_t}
 +\sum_{1\le t<u\le |J|}s_{j_tj_u}.
 \end{gathered}
 \label{eq:ws-kn-scaling}
\end{equation}
Here $\mathrm{KN}_I$ and $\mathrm{KN}_J$ contain
the within-block factors and distances to $b$,
and depend only on the shape variables.
The vertex at infinity is understood through
its normalized limit.
On the zero-integer rectangle, the cross product
in Eq.~\eqref{eq:ws-kn-scaling} is one:
the KN dependence separates into two shape
factors and a single power of the relative scale.

For general integer entries, the cross factors
must be combined with the remaining correlator.
Assume that, after separating the noninteger KN factors,
the complete gauge-fixed correlator admits a finite
rational representation with poles only at
external-puncture collisions.
Equation~\eqref{eq:ws-cross-regularity} then cancels
the cross denominators in the combined expression,
including negative integer KN powers.
The remaining cross dependence is polynomial,
while all other factors scale homogeneously.
Regular cross interactions can survive, but contribute
only finitely many homogeneous sectors.
Related polynomial and scaleless structures appear
in curve-integral descriptions of string zeros
\cite{Arkani-Hamed:2023swr,Chang:2025cqe}.

The coordinate Jacobian contributes
$\lambda^{|J|-1}$.
Including the complete gauge-fixed measure gives
\begin{equation}
 \begin{gathered}
 \Omega_{\rm rect}^{\rm open}
 =\frac{d\lambda}{\lambda}
 \sum_\ell\lambda^{\Delta_\ell}
 \widehat\Omega_\ell(x,y),\\
 \Delta_\ell=|J|+\beta_J+r_\ell,
 \qquad r_\ell\in\mathbb Z.
 \end{gathered}
 \label{eq:ws-scale-decomposition}
\end{equation}
The sum is finite, and the $\lambda$-independent
forms $\widehat\Omega_\ell$ retain all shape variables
and external-state data.
Each sector carries the formally scaleless integral
$\int_0^\infty d\lambda\,\lambda^{\Delta_\ell-1}$\cite{Arkani-Hamed:2023swr}.

On the sphere, use the same algebraic coordinate
change with complex $x,y$, no ordering inequalities,
and $\lambda\in\mathbb C^\times$.
The Jacobian is now $|\lambda|^{2(|J|-1)}$.
Under complete chiral cross regularity, the finite
decomposition becomes
\begin{equation}
 \Omega_{\rm rect}^{\rm closed}
 =\frac{d^2\lambda}{|\lambda|^2}
 \sum_\ell
 \lambda^{h_\ell}\bar\lambda^{\widetilde h_\ell}
 \widehat\Omega_\ell.
 \label{eq:ws-closed-scale-decomposition}
\end{equation}
The shape forms are again independent of $\lambda$.
Writing $\lambda=Re^{i\theta}$ separates the angular
weight $h_\ell-\widetilde h_\ell$ from the radial
weight $h_\ell+\widetilde h_\ell$.
Single-valuedness makes the angular weight an integer.
Angular integration removes sectors with nonzero
angular weight, while the surviving sectors carry
scaleless radial integrals.
External spins and mass levels enter the scale weights
and shape forms without changing this mechanism.

This finite decomposition gives a relative-scale
picture of the zeros.
An independent period proof would additionally require
matching the scaleless prescription to analytic
continuation of the complete disk and sphere periods
\cite{Mizera:2017cqs,Brown:2019wna},
including overlapping scale and shape degenerations.
The amplitude-level results follow from monodromy
and KLT.
Local OPE calculations, relative-scale measures
and illustrative four-point Euler periods
are given in the SM.

\medskip
\sec{Discussion}

Integer rectangles first highlighted by colored-scalar
hidden zeros organize a more general behavior of string
amplitudes. With arbitrary physical external wavefunctions,
the surviving color-ordered amplitude admits zero-locus
factorization: a weighted sum of on-shell
$3\times(n-1)$ products, under the stated regularity and
invertibility conditions. Hidden zeros follow when the
external wavefunctions make all contributing three-point
functionals vanish.
KLT extends this physical fusion criterion to uncolored
closed-string states, with the required persistence of
the chiral conditions. The cross OPEs give a local
description of these fusion conditions across external
species and finite mass levels. Under stronger cross-OPE
regularity, a finite relative-scale decomposition reveals
formally scaleless sectors.

The zero-locus factorization formula provides finite-amplitude data
on special kinematic loci, suggesting further constraints
on amplitude reconstruction.
Connections to splitting
\cite{Cachazo:2021wsz,Cao:2024gln,Cao:2024qpp,
Arkani-Hamed:2024fyd,Li:2026cwa}
and diagrammatic and expansion-based descriptions
\cite{Li:2024bwq,Zhou:2024ddy,Huang:2025blb}
offer complementary perspectives on this question.
Further connections include reconstruction from hidden zeros
\cite{Rodina:2024yfc,Zhou:2026isc},
on-shell derivations of zeros and splitting
\cite{Jones:2025rbv,Feng:2025ofq},
recursive determination of boundary terms
\cite{Zhou:2026zab},
and string Britto--Cachazo--Feng--Witten recursion
\cite{Cheung:2010vn,Boels:2010bv}.
It would be interesting to determine whether these
restrictions, combined with physical factorization
and suitable large-momentum behavior, remove ambiguities
left by pole residues alone.

A complementary question is how the physical cubic-zero
conditions are encoded in other amplitude representations.
Scalar scaffolding \cite{Arkani-Hamed:2023jry},
curve integrals \cite{Chang:2025cqe},
disk-integral decompositions
\cite{Mafra:2011nv,Mafra:2011nw,Huang:2016tag,He:2018pol},
and bosonic building blocks for superstring amplitudes
\cite{Cao:2025lzv}
provide concrete settings for tracing these conditions
through fusion residues and cancellations among
kinematic factors.
Explicit massive couplings
\cite{Boels:2012if,Markou:2023ffh}
supply further state-dependent tests.
Comparisons with massive field-theory zeros
\cite{CarrilloGonzalez:2026lnu},
higher-derivative zeros \cite{Zhou:2025tvq}
and single-valued disk--sphere relations
\cite{Stieberger:2013wea,Brown:2019wna}
could clarify the low-energy expansion and the relation
between open- and closed-string zero conditions.
They could also reveal when the fusion-zero
criterion is manifest in individual building blocks
and when it emerges only after their combination.

\medskip
\sec{Acknowledgements}

We thank Qu Cao for helpful discussions.
OpenAI's GPT-6 Astra model was used under the authors' direction to assist with derivations, computational checks and manuscript preparation.
The authors reviewed and verified all AI-assisted material
and scientific conclusions and take full responsibility
for the content.  The work
of S.H. is supported by the National Natural Science Foundation of China under
Grant Nos.~12225510 and 12247103, and by the New Cornerstone Science
Foundation. Y.Z.'s work is supported by the National Natural Science
Foundation of China under Grant No.~12405086 and by the Ningbo Science \& Technology Bureau under program grant No.~2024DJRC010009.
The work of F.Z. is supported in part by the Science Challenge Project (No. TZ2025012), and NSAF No. U2330401.

\bibliographystyle{apsrev4-2}

\bibliography{Refs}

\onecolumngrid
\clearpage

\begin{center}
{\large\bfseries SUPPLEMENTAL MATERIAL}
\end{center}

\vspace{1.5em}

\appendix

\section{Conventions and physical sewing}
\label{app:conventions}

We fix the normalization, physical pairing and
fermion-ordering conventions entering
Eq.~\eqref{eq:sz-physical-sewing}.
External spinor wavefunctions are commuting analytic
variables; statistical signs are recorded explicitly.

\subsection{Normalization and physical pairing}
\label{app:normalization}

We use a mostly-plus metric, outgoing momenta and
\begin{equation}
 \langle X^\mu(z)X^\nu(w)\rangle
 =-2\alpha'\eta^{\mu\nu}\log(z-w),
 \qquad
 s_{rs}=2\alpha'k_r\cdot k_s.
 \label{eq:app-a-x-ope}
\end{equation}
Ordered disk periods have positive integration
density and a fixed Faddeev--Popov orientation.
A common coupling and the momentum-conserving
delta function are stripped.

Let $\mathcal A_n$ denote the raw ghost-fixed
vertex integral. Our amplitudes and kinetic
pairing are
\begin{equation}
 A_n=N_n\mathcal A_n,
 \qquad
 N_n=-\frac{i^n}{2\alpha'^2},
 \qquad
 G_{\rm kin}=\frac{G^{\rm raw}}{2\alpha'},
 \label{eq:app-a-amplitude-normalization}
\end{equation}
where $G^{\rm raw}$ is the complete pairing on
physical BRST cohomology, including the required
ghost and picture insertions.
For the fusion of legs $i,j$, write $K_{ij}=k_i+k_j$.
The channel propagator is
$(G_{\rm kin}^{-1})_{\mathsf b\mathsf a}
/(K_{ij}^2+M_{K_{ij}}^2)$.
The pairing is evaluated with the lower-amplitude
state $K_{ij}^{\mathsf a}$ first and the cubic dual
$-K_{ij}^{\mathsf b}$ second; its inverse contracts
the cubic dual with the lower-amplitude state
in that order.

For the raw residue in the variable $\delta_{ij}$,
the conversion to normalized physical sewing is
\begin{equation}
 \frac{N_n}{N_3N_{n-1}}(G^{\rm raw})^{-1}
 =
 \frac{2\alpha'^2}{2\alpha'}G_{\rm kin}^{-1}
 =
 \alpha'G_{\rm kin}^{-1}.
 \label{eq:app-a-normalization-ratio}
\end{equation}
This gives the coefficient $\alpha'$ in
Eq.~\eqref{eq:sz-physical-sewing}, consistently
with $\delta_{ij}=\alpha'(K_{ij}^2+M_{K_{ij}}^2)$.

For the bosonic representatives
$V_T=:e^{ik\cdot X}:$ and
$V_g=:\epsilon\cdot\partial X\,e^{ik\cdot X}:$,
the pairings used in the low-point calibrations are
\begin{equation}
 G_T^{\rm raw}=1,
 \qquad
 G_g^{\rm raw}=-2\alpha'\eta\big|_{\rm phys},
 \label{eq:app-a-bosonic-metrics}
\end{equation}
with the vector pairing taken on transverse
polarizations modulo gauge equivalence.

\label{app:nsr-sewing}
For superstring channels, the physical pairing uses
complementary pictures of total picture $-2$
\cite{Friedan:1985ge}.
We use pictures $(-1,-1)$ in the Neveu--Schwarz (NS)
sector, and pair a lower Ramond state of picture
$-1/2$ with a cubic dual of picture $-3/2$.
At nonzero momentum, other picture representatives
can be used by transferring both the physical states
and their pairing
\cite{Belopolsky:1997jz,Berkovits:1997mc}.
We follow the physical sewing prescriptions of
Refs.~\cite{Kunitomo:2015usa,Kunitomo:2016bhc}.

\subsection{Fermion ordering and oriented residues}
\label{app:fermion-reference}
\label{app:oriented-sewing}

Fix a reference order $\mathfrak f$ of the labelled
external fermions. If $F(w)$ is their subsequence
in a word $w$, define
\begin{equation}
 A_w^{\rm ref}=\chi_F(w)C_w,
 \qquad
 \chi_F(w)=
 \operatorname{sgn}\bigl(\mathfrak f\longrightarrow F(w)\bigr),
 \label{eq:app-a-reference-amplitude}
\end{equation}
where $C_w$ has fermionic slots ordered as in $w$.
For an even total fermion number, the reference
amplitudes obey ordinary cyclicity.
The main text suppresses the superscript ``ref''.

Each sewn child inherits the reference subsequence
of its surviving external fermions and appends
its internal fermion label last, when present.
If $\chi_3,\chi_{n-1}$ convert these child reference
orders to their displayed words, the conversion
factor for the parent $P_t^{(i,\rho)}$ is
\begin{equation}
 \sigma_{ij_t}^{\rho}
 =
 \chi_F(P_t^{(i,\rho)})\,\chi_3\,\chi_{n-1}.
 \label{eq:app-a-sewing-permutation}
\end{equation}
For bosonic states all three signs equal one.

The cubic reversal sign follows from three-point
physical disk monodromy
\cite{Bjerrum-Bohr:2009ulz,Stieberger:2009hq,Boels:2014dka}.
At the selected fusion level, $s_{ij}=q_{ij}$,
and in a common fermion reference order,
\begin{equation}
 A_3^{\rm ref}(i,j,-K_{ij})
 +e^{i\pi q_{ij}}A_3^{\rm ref}(j,i,-K_{ij})=0.
 \label{eq:app-a-three-point-monodromy}
\end{equation}
The same physical dual state and pairing are
used in both terms.
Converting to component amplitudes contributes
the permutation sign $(-1)^{\mathfrak p_i\mathfrak p_j}$,
giving
\begin{equation}
 C_3(j,i,-K_{ij})=\tau_{ij}C_3(i,j,-K_{ij}),
 \qquad
 \tau_{ij}=-(-1)^{q_{ij}+\mathfrak p_i\mathfrak p_j},
 \label{eq:app-a-cubic-reversal}
\end{equation}
where $\mathfrak p_i=0$ for bosonic or NS states
and $\mathfrak p_i=1$ for Ramond states.
In $P_t^{(i,\rho)}$, the colliding pair appears
in the order $(j_t,i)$, whereas
Eq.~\eqref{eq:sz-physical-sewing} uses $(i,j_t)$.
Their reversal therefore supplies $\tau_{ij_t}$.
Together with
Eqs.~\eqref{eq:app-a-normalization-ratio}
and \eqref{eq:app-a-sewing-permutation},
this gives the oriented physical sewing
in Eq.~\eqref{eq:sz-physical-sewing}.

For neighboring auxiliary words, exchanging
the adjacent external labels changes the parent
reference sign by $(-1)^{\mathfrak p_i\mathfrak p_{j_t}}$.
Their complete lower amplitude is identical.
Consequently,
\begin{equation}
 \operatorname*{Res}_{\delta_{ij_t}=0}P_{t-1}
 =
 (-1)^{\mathfrak p_i\mathfrak p_{j_t}}\tau_{ij_t}
 \operatorname*{Res}_{\delta_{ij_t}=0}P_t
 =
 -(-1)^{q_{ij_t}}
 \operatorname*{Res}_{\delta_{ij_t}=0}P_t.
 \label{eq:app-a-neighboring-residues}
\end{equation}
This relation holds before restricting the other
cross invariants and gives the species-independent
residue cancellation used in the rectangular limit.
The resulting sign is illustrated by the
four-fermion relation in
Appendix~\ref{app:massive-ramond-calibration}.

\section{Monodromy and the joint rectangular limit}
\label{app:insertion-proof}
\label{app:joint-pole-subtraction}

We spell out the cancellation of neighboring fusion
poles used in the main-text proof.

Momentum conservation gives
\begin{equation}
 \sum_{s\ne i}s_{is}=2\alpha'M_i^2\in2\mathbb Z,
 \label{eq:app-b-even-mass-shift}
\end{equation}
both for bosonic levels $\alpha'M_i^2=L_i-1$
and NS and Ramond levels $\alpha'M_i^2=\ell_i$ after the
Gliozzi--Scherk--Olive (GSO) projection.
These even mass-shell shifts leave the sine
coefficients unchanged, giving
Eq.~\eqref{eq:sz-full-monodromy} from disk monodromy
\cite{Bjerrum-Bohr:2009ulz,Stieberger:2009hq,
Boels:2010bv,Ma:2011um}.

Fix a row $(i,\rho)$ and suppress its labels.
Using the auxiliary words $P_t$ of the main text,
define
\begin{equation}
 \begin{gathered}
 d_{i,t}=C_{i,t}-c_{i,t}
 =\sum_{u=1}^{t}\delta_{ij_u},\qquad
 \widetilde B_t=\sum_{u=t}^{|J|}(-1)^{c_{i,u}}P_u,\\
 d_{i,0}=c_{i,0}=0.
 \end{gathered}
 \label{eq:sz-rephased-tails}
\end{equation}
Since $c_{i,t}$ is integer, summation by parts gives
\begin{equation}
 \sum_{t=1}^{|J|}\sin(\pi C_{i,t})P_t
 =
 \sum_{t=1}^{|J|}
 \bigl[\sin(\pi d_{i,t})-\sin(\pi d_{i,t-1})\bigr]
 \widetilde B_t.
 \label{eq:sz-tail-summation}
\end{equation}

The moving leg in $P_u$ can have selected fusion
poles only at its two neighbors in $J$.
For $v>t$, the two contributions to the
$\delta_{ij_v}$ residue of $\widetilde B_t$ cancel:
\begin{equation}
 \operatorname*{Res}_{\delta_{ij_v}=0}\widetilde B_t
 =
 (-1)^{c_{i,v-1}}
 \operatorname*{Res}_{\delta_{ij_v}=0}P_{v-1}
 +(-1)^{c_{i,v}}
 \operatorname*{Res}_{\delta_{ij_v}=0}P_v
 =0.
 \label{eq:app-b-tail-cancellation}
\end{equation}
Here we used
Eq.~\eqref{eq:app-a-neighboring-residues} and
$c_{i,v}=c_{i,v-1}+q_{ij_v}$.
The cancellation holds before restricting the
other cross invariants.

Under the assumptions of the main text, the tail
therefore has only the possible simple pole
$\delta_{ij_t}=0$, supplied by $P_t$.
We can write
\begin{equation}
 \widetilde B_t
 =\frac{(-1)^{c_{i,t}}r_t}{\delta_{ij_t}}+H_t,
 \qquad
 r_t,H_t\ \text{regular near the rectangle},
 \qquad
 r_t|_{h_m(\mathbf q)}=R_{ij_t}^{\rho}.
 \label{eq:sz-joint-tail}
\end{equation}
For a nonspectral pair, $r_t$ can be chosen to vanish.

Since $d_{i,t}=d_{i,t-1}+\delta_{ij_t}$,
\begin{equation}
 \frac{\sin(\pi d_{i,t})-\sin(\pi d_{i,t-1})}
 {\delta_{ij_t}}
 =\pi\int_0^1
 \cos\!\left[\pi(d_{i,t-1}+v\delta_{ij_t})\right]\,dv
 \longrightarrow\pi.
 \label{eq:app-b-sine-quotient-integral}
\end{equation}
Substituting Eq.~\eqref{eq:sz-joint-tail} into
Eq.~\eqref{eq:sz-tail-summation}, each pole term
contributes $\pi(-1)^{c_{i,t}}R_{ij_t}^{\rho}$.
The regular remainders multiply sine differences
that vanish on the rectangle.
Thus
\begin{equation}
 \left.
 \sum_{t=1}^{|J|}\sin(\pi C_{i,t})P_t
 \right|_{h_m(\mathbf q)}
 =
 \pi\sum_{t=1}^{|J|}(-1)^{c_{i,t}}R_{ij_t}^{\rho},
 \label{eq:app-b-rectangular-source}
\end{equation}
independently of the relative approach rates.
Together with the left-hand side of the monodromy
relation, this gives
Eq.~\eqref{eq:sz-residue-system}.

The insertion determinant is not identically
zero in unconstrained invariant space.
Order the columns lexicographically and pair
row $(i,\rho)$ with the column whose word starts
with $i$ and continues with $\rho$.
Specialize
\[
 s_{ib}=\mathrm iT,\qquad
 s_{ih}=\mathrm iT
 \quad(i,h\in I,\ i\ne h),\qquad T>0.
\]
Inserting $i$ after $r$ entries of $\rho$ gives
the supported row entry
$\mathrm i\sinh[\pi T(m-r)]$, $r=0,\ldots,m-1$.
The diagonal corresponds to $r=0$ and, for
sufficiently large $T$,
\[
 \sinh(m\pi T)>
 \sum_{k=1}^{m-1}\sinh(k\pi T).
\]
Strict diagonal dominance proves invertibility;
the case $m=1$ is immediate.

Formal nonvanishing does not establish
invertibility on every physical Gram component.
An invariant assignment must admit an on-shell,
momentum-conserving realization whose Gram rank
does not exceed the actual spacetime dimension.
The inverse representation is used only on
physical families where the restricted
determinant is nonzero.

\section{External-state conditions for fusion zeros}
\label{app:fusion-zero-conditions}

We give explicit external-wavefunction conditions
for the selected physical cubic couplings to vanish:
\begin{equation}
 A_3(i,j,-K_{ij}^{\eta})=0
 \qquad\forall\eta\in\mathcal H_{ij}^{\rm phys}(q_{ij}).
 \label{eq:app-c-zero-functional}
\end{equation}
The conditions apply to spectral cross pairs and
must hold simultaneously across the rectangle,
with nonzero physical external states.
Local OPE calculations and complete amplitude examples
are given in Appendices~\ref{app:ope-scale}
and~\ref{app:examples}, respectively.

\subsection{Massless external states}
\label{app:fusion-zero-massless}

For two gluons at $q_{ij}=0$, the bridges
\begin{equation}
 \epsilon_i\cdot k_j=k_i\cdot\epsilon_j
 =\epsilon_i\cdot\epsilon_j=0
 \label{eq:app-c-zero-gluon-bridges}
\end{equation}
annihilate the complete massless-vector cubic
functional in both the bosonic string and the
superstring, including the finite-$\alpha'$
bosonic contribution
\cite{Sagnotti:2010at,Schlotterer:2010kk}.
For two massless bosonic-string gluons at
$q_{ij}=1$, the selected state is the tachyon,
whose cubic coupling vanishes when
\begin{equation}
 \epsilon_i\cdot\epsilon_j
 -2\alpha'(\epsilon_i\cdot k_j)(k_i\cdot\epsilon_j)=0.
 \label{eq:app-c-zero-tachyon-coupling}
\end{equation}
Thus Eq.~\eqref{eq:app-c-zero-gluon-bridges}
also removes this channel.

For massless Ramond states, use commuting
ten-dimensional GSO-projected wavefunctions and define
\begin{equation}
 \slashed k_i\chi_i=0,\qquad
 \bar\chi_i=\chi_i^TC,\qquad
 \slashed f_i=\frac14 f_{i,\mu\nu}\Gamma^{\mu\nu},
 \label{eq:app-c-zero-field-strength}
\end{equation}
where $f_i^{\mu\nu}=k_i^\mu\epsilon_i^\nu-k_i^\nu\epsilon_i^\mu$
and $\Gamma^{\mu\nu}=\frac12[\Gamma^\mu,\Gamma^\nu]$.
In the table below, $g_i$ and $f_i$ label gluon and
Ramond states, respectively; $f_i^{\mu\nu}$ denotes
the gluon field strength.
At $q_{ij}=0$, sufficient fermionic bridges are
\cite{Zhang:2026wye}
\begin{equation}
 \begin{array}{c|l}
 \text{cross pair}&\text{sufficient conditions}\\ \hline
 f_i f_j&
 \bar\chi_i\Gamma^\mu\chi_j=0
 \quad\text{for every }\mu
 \\[2pt]
 g_i f_j&
 \epsilon_i\cdot k_j=0,\qquad
 \slashed f_i\chi_j=0
 \\[2pt]
 f_i g_j&
 k_i\cdot\epsilon_j=0,\qquad
 \bar\chi_i\slashed f_j=0
 \end{array}
 \label{eq:app-c-zero-fermionic-bridges}
\end{equation}
These bridges set the corresponding vector current
or spinor source to zero.
Under dimensional reduction, the first row
includes the internal vector components whenever
the corresponding physical scalars are retained.

For massless bosonic-string gluons, $q_{ij}\ge2$
is nonspectral. For massless GSO-projected inputs,
every positive integer is nonspectral.
Negative integers select massive intermediate states,
whose cubic couplings must be examined separately.
Negative-integer bosonic gluon fusion is discussed
in Appendix~\ref{app:four-point-calibration}.
The role of the physical spectrum in integer roots
is discussed in Ref.~\cite{Boels:2014dka}.

\subsection{Massive bosonic external states}
\label{app:fusion-zero-massive}

Consider bosonic leading-Regge states at oscillator
levels $L_i,L_j$, represented by symmetric,
transverse and traceless tensors $E_i,E_j$ and their
standard vertex operators \cite{Sagnotti:2010at}, with
\begin{equation}
 \alpha'M_i^2=L_i-1.
 \label{eq:app-c-zero-regge-mass}
\end{equation}
Impose the momentum bridges
\begin{equation}
 E_i\cdot k_j=0,\qquad E_j\cdot k_i=0,
 \label{eq:app-c-zero-tensor-momentum}
\end{equation}
where contraction in any tensor slot vanishes.
Define the partial cross contractions
\begin{equation}
 (C_r)_{AB}
 =E_{i,A\mu_1\cdots\mu_r}
 E_{j,B}{}^{\mu_1\cdots\mu_r},
 \qquad1\le r\le\min(L_i,L_j),
 \label{eq:app-c-zero-partial-contractions}
\end{equation}
where $A,B$ denote the remaining Lorentz multi-indices.

For $q_{ij}\ge0$, a sufficient tensor condition is
\begin{equation}
 C_{r_0}=0,\qquad
 r_0=\left\lfloor\frac{q_{ij}}2\right\rfloor+1.
 \label{eq:app-c-zero-regge-rule}
\end{equation}
If $r_0>\min(L_i,L_j)$, the momentum bridges alone
suffice. The OPE derivation is given in
Appendix~\ref{app:massive-tensor-ope}.
The selected fusion level is
\begin{equation}
 L_K=L_i+L_j-1-q_{ij}.
 \label{eq:app-c-zero-regge-fusion}
\end{equation}
For two first-massive tensors, $L_i=L_j=2$,
write $E_i:E_j:=E_{i,\mu\nu}E_j^{\mu\nu}$.
The spectral rows below assume the momentum bridges
in Eq.~\eqref{eq:app-c-zero-tensor-momentum}
and list the additional tensor conditions:
\begin{equation}
 \begin{array}{c|c|c}
 q_{ij}&L_K&\text{additional tensor condition}\\ \hline
 0&3&C_1=0\\
 1&2&C_1=0\\
 2&1&C_2=E_i:E_j=0\\
 3&0&C_2=E_i:E_j=0\\
 \ge4&\text{nonspectral}&\text{none}
 \end{array}.
 \label{eq:app-c-zero-first-massive-table}
\end{equation}
The nonspectral row requires no state condition.
Under the momentum bridges, it suffices to impose
$E_i:E_j=0$ at $q_{ij}=2,3$; the partial contraction
$C_1$ may remain nonzero.
This gives Eq.~\eqref{eq:hz-first-massive-tensors}.
The corresponding four-point amplitude relations
are given in Appendix~\ref{app:massive-four-point}.

\subsection{Massive Ramond external states}
\label{app:fusion-zero-massive-ramond}

Consider a massless Ramond state $f_i$ and a
leading-Regge Ramond state $F_j^{(\ell)}$, with
\begin{equation}
 k_i^2=0,\qquad
 k_j^2=-\frac{\ell}{\alpha'},\qquad\ell\ge1.
 \label{eq:app-c-zero-ramond-masses}
\end{equation}
The independent chiral tensor-spinor
$v_j^{\mu_1\cdots\mu_\ell}$ is symmetric,
transverse and gamma-traceless.
The complementary component of the physical
vertex is fixed by the massive Dirac relation.

At $q_{ij}=\ell$, the selected fusion space
is the massless NS vector space. Define
\begin{equation}
 W_j^\mu
 =v_j^{\mu\nu_1\cdots\nu_{\ell-1}}
 k_{i,\nu_1}\cdots k_{i,\nu_{\ell-1}}.
 \label{eq:app-c-zero-ramond-contractions}
\end{equation}
Specializing the complete cubic coupling of
Ref.~\cite{Schlotterer:2010kk}, Eq.~(3.14), gives,
up to a nonzero normalization,
\begin{equation}
 A_3(f_i,F_j^{(\ell)},-K_{ij,g}^{\eta})
 \propto\epsilon^\eta_{K_{ij},\mu}\mathcal J_{ij}^{\mu},
 \qquad
 \mathcal J_{ij}^{\mu}
 =\bar\chi_i\slashed k_jW_j^\mu
 -\bar\chi_i\Gamma^\mu(k_i\cdot W_j).
 \label{eq:app-c-zero-ramond-current}
\end{equation}
Both terms are required by the complete physical
Ramond vertex.
Transversality of $W_j^\mu$ to $k_j$ and the
massless Dirac equation give the Ward identity
\begin{equation}
 K_{ij,\mu}\mathcal J_{ij}^{\mu}
 =\bar\chi_i(\slashed k_j-\slashed K_{ij})(k_i\cdot W_j)
 =-\bar\chi_i\slashed k_i(k_i\cdot W_j)=0.
 \label{eq:app-c-zero-ramond-ward}
\end{equation}

Setting $\mathcal J_{ij}^{\mu}=0$ is sufficient.
A convenient stronger choice is
\begin{equation}
 k_i\cdot W_j=0,\qquad
 \bar\chi_i\slashed k_jW_j^\mu=0
 \quad\text{for every }\mu.
 \label{eq:app-c-zero-ramond-bridges}
\end{equation}
For the first-massive vector-spinor, $\ell=1$,
one has $W_j^\mu=v_j^\mu$.
Equation~\eqref{eq:app-c-zero-ramond-current} then
reduces to Eq.~\eqref{eq:hz-first-massive-ramond-current},
and the sufficient conditions become
Eq.~\eqref{eq:hz-first-massive-ramond} at $q_{ij}=1$.

\section{Examples of zero-locus factorization and hidden zeros}
\label{app:examples}

We give explicit four- and six-point amplitude
relations and their fusion-zero conditions.
All relations use the domain specified in the main
text, with nonzero insertion denominators.
Each state sum runs over the complete physical
fusion space of the indicated channel.

\subsection{Four gluons}
\label{app:four-point-calibration}

For four massless bosonic-string gluons, take
\begin{equation}
 s_{13}=q\in\mathbb Z,\qquad s_{12}+s_{23}+q=0,
 \qquad
 K_{13}=k_1+k_3,\qquad
 M_{K_{13}}^2=-\frac{q}{\alpha'}.
 \label{eq:app-c-four-invariants}
\end{equation}
With $I=\{1\}$, $b=2$, $J=(3)$ and $a=4$,
zero-locus factorization gives
\begin{equation}
 \left.A_4(1_g,2_g,3_g,4_g)\right|_{s_{13}=q}
 =-\frac{\pi\alpha'}{\sin(\pi s_{12})}
 \sum_{\mathsf a,\mathsf b}
 A_3(1_g,3_g,-K_{13}^{\mathsf b})\,
 (G_{\rm kin}^{-1})_{\mathsf b\mathsf a}\,
 A_3(2_g,K_{13}^{\mathsf a},4_g).
 \label{eq:app-c-four-factorization}
\end{equation}
The integer assignments fall into four classes:
\begin{equation}
 \begin{array}{c|c|c}
 q&L_{K_{13}}&\text{physical fusion space}\\ \hline
 q\ge2&\text{---}&\text{nonspectral}\\
 1&0&\text{tachyon}\\
 0&1&\text{massless vector}\\
 q\le-1&1-q&\text{massive states}
 \end{array}.
 \label{eq:app-c-four-fusion-spectrum}
\end{equation}

For $q\ge2$, the fusion space is empty.
The amplitude therefore vanishes without
additional polarization conditions:
\begin{equation}
 \left.A_4(1_g,2_g,3_g,4_g)\right|_{s_{13}=q}=0,
 \qquad q\ge2.
 \label{eq:app-c-four-nonspectral-zero}
\end{equation}

At $q=1$, the selected state is the tachyon.
Its cubic coupling vanishes under
\begin{equation}
 \epsilon_1\cdot\epsilon_3
 -2\alpha'(\epsilon_1\cdot k_3)
 (k_1\cdot\epsilon_3)=0,
 \label{eq:app-c-four-tachyon-condition}
\end{equation}
which sets the four-point amplitude to zero.

At $q=0$, the selected states are massless vectors.
The gluon bridges
\begin{equation}
 \epsilon_1\cdot k_3=k_1\cdot\epsilon_3
 =\epsilon_1\cdot\epsilon_3=0
 \label{eq:app-c-four-zero-bridges}
\end{equation}
annihilate their complete cubic functional, giving
\begin{equation}
 \left.A_4(1_g,2_g,3_g,4_g)\right|_{s_{13}=0}=0.
 \label{eq:app-c-four-massless-zero}
\end{equation}
These bridges also satisfy the tachyon condition above.

For $q\le-1$, the selected fusion space contains
the leading-Regge state at level $L=1-q\ge2$.
In the critical bosonic string, set
$K=K_{13}$ and $p=k_1-k_3$, so that
$p^2=-q/\alpha'\ne0$. Since $k_1\cdot k_3\ne0$,
gauge transformations put both gluon polarizations
in $K^\perp\cap p^\perp$.
For a physical leading-Regge polarization
$E=z^{\otimes L}$ with a complex auxiliary vector
$z$ satisfying $z\cdot K=z^2=0$,
Eq.~(4.6) of Ref.~\cite{Sagnotti:2010at} reduces,
up to a nonzero overall factor, to
\[
 A_3\propto(z\cdot p)^{L-2}
 \left[L(L-1)(z\cdot\epsilon_1)(z\cdot\epsilon_3)
 +\frac{\alpha'}2(\epsilon_1\cdot\epsilon_3)
 (z\cdot p)^2\right].
\]
If the coupling vanished for every physical $E$,
the quadratic expression in brackets would vanish
on the dense subset $z\cdot p\ne0$ of the irreducible
null cone in the nondegenerate, 25-dimensional
space $K^\perp$, and hence on the entire cone.
It would therefore be proportional to $z^2$.
Its matrix has rank at most three, whereas the
metric on $K^\perp$ has rank 25, so the quadratic
expression must vanish identically.
Restricting it to $K^\perp\cap p^\perp$ then gives
$(z\cdot\epsilon_1)(z\cdot\epsilon_3)\equiv0$,
forcing one common-transverse polarization to vanish.
The corresponding external gluon is pure gauge.
This rules out the pairwise cubic-zero mechanism for two
nonzero physical gluons; it does not exclude amplitude zeros
produced by cancellations among nonzero fusion contributions.

\subsection{Massive external tensors}
\label{app:massive-four-point}

Take first-massive bosonic tensors $1_M,3_M$
and massless gluons $2_g,4_g$, with
\begin{equation}
 M_1^2=M_3^2=\frac1{\alpha'},\qquad
 s_{13}=q,\qquad s_{12}+s_{23}=2-q.
 \label{eq:app-c-massive-invariants}
\end{equation}
The selected momentum is $K_{13}=k_1+k_3$,
with mass $M_{K_{13}}^2=(2-q)/\alpha'$.
The amplitude relation reads
\begin{equation}
 \left.A_4(1_M,2_g,3_M,4_g)\right|_{s_{13}=q}
 =-\frac{\pi\alpha'}{\sin(\pi s_{12})}
 \sum_{\mathsf a,\mathsf b}
 A_3(1_M,3_M,-K_{13}^{\mathsf b})\,
 (G_{\rm kin}^{-1})_{\mathsf b\mathsf a}\,
 A_3(2_g,K_{13}^{\mathsf a},4_g).
 \label{eq:app-c-massive-factorization}
\end{equation}
At $q=2,3$, the selected states are the massless
vector and tachyon, respectively.
The tensor conditions
\begin{equation}
 E_1\cdot k_3=E_3\cdot k_1=0,\qquad E_1:E_3=0
 \label{eq:app-c-massive-zero-conditions}
\end{equation}
annihilate both selected cubic functionals,
as described in Appendix~\ref{app:fusion-zero-massive}.
Thus
\begin{equation}
 \left.A_4(1_M,2_g,3_M,4_g)\right|_{s_{13}=2}
 =
 \left.A_4(1_M,2_g,3_M,4_g)\right|_{s_{13}=3}
 =0.
 \label{eq:app-c-massive-zeros}
\end{equation}
No additional conditions on the gluon
polarizations are needed.

For an explicit realization, choose orthogonal unit
vectors $u,v$ transverse
to all momenta and set
\[
 E_1=u\otimes u-v\otimes v,\qquad
 E_3=u\otimes v+v\otimes u.
\]
The tensors are transverse and traceless, with
\[
 E_1:E_3=0,\qquad
 (C_1)_{\mu\nu}=u_\mu v_\nu-v_\mu u_\nu\ne0.
\]

\subsection{A massive Ramond external state}
\label{app:massive-ramond-calibration}

Consider three massless Ramond states $f_1,f_2,f_3$
and a first-massive Ramond state $F_4$ in ten
dimensions, with $M_{F_4}^2=1/\alpha'$.
Momentum conservation gives
\begin{equation}
 s_{12}+s_{13}+s_{14}=0.
 \label{eq:app-c-massive-ramond-invariants}
\end{equation}
On $s_{13}=0$, the pair $13$ selects a massless
NS vector with momentum $K_{13}=k_1+k_3$.
For fermion reference order $(1,2,3,4)$,
\begin{equation}
 \left.A_4(f_1,f_2,f_3,F_4)\right|_{s_{13}=0}
 =\frac{\pi\alpha'}{\sin(\pi s_{12})}
 \sum_{\mathsf a,\mathsf b}
 A_3(f_1,f_3,-K_{13,g}^{\mathsf b})\,
 (G_{\rm kin}^{-1})_{\mathsf b\mathsf a}\,
 A_3(f_2,K_{13,g}^{\mathsf a},F_4).
 \label{eq:app-c-massive-ramond-factorization}
\end{equation}
The positive coefficient follows from
$\tau_{13}=\sigma_{13}=1$.
The massless spinor bridge
\begin{equation}
 \bar\chi_1\Gamma^\mu\chi_3=0
 \qquad\text{for every }\mu
 \label{eq:app-c-massive-ramond-zero-condition}
\end{equation}
removes the $13$ cubic coupling and gives
\begin{equation}
 \left.A_4(f_1,f_2,f_3,F_4)\right|_{s_{13}=0}=0.
 \label{eq:app-c-massive-ramond-zero}
\end{equation}
Alternatively, the massive Ramond condition can remove
the other three-point factor in
Eq.~\eqref{eq:app-c-massive-ramond-factorization}.
Momentum conservation gives
$s_{24}=s_{13}+1$, so $s_{13}=0$ also puts the
$24$ fusion channel at its massless-vector value
$s_{24}=1$. Define
\[
 j_{13}^{\mu}=\bar\chi_1\Gamma^\mu\chi_3,
 \qquad
 \mathcal J_{24}^{\mu}
 =\bar\chi_2\slashed k_4v_4^\mu
 -\bar\chi_2\Gamma^\mu(k_2\cdot v_4).
\]
Here $\mathcal J_{24}$ is the $\ell=1$ current in
Eq.~\eqref{eq:app-c-zero-ramond-current}.
The condition $\mathcal J_{24}^{\mu}=0$ for every
$\mu$ annihilates this factor and therefore also gives
Eq.~\eqref{eq:app-c-massive-ramond-zero}.

The complete four-fermion amplitude in
Ref.~\cite{Schlotterer:2010kk}, Eq.~(4.24), gives
an independent check. At the first massive level,
the variables of that reference are
$s=s_{12}$, $t=s_{13}$ and $u=s_{14}-1$.
Only the $13$-fusion structure survives at $s_{13}=0$,
where $s_{14}=-s_{12}$ and its beta factor is
\[
 \left.B(s_{12}+1,s_{14})\right|_{s_{13}=0}
 =-\frac{\pi}{\sin(\pi s_{12})}.
\]
With the spinor conventions matched, its numerator
is proportional to
\[
 -j_{13,\mu}\mathcal J_{24}^{\mu}.
\]
The overall factor depends on the external-vertex
normalization.
Both currents are transverse to $K_{13}=-K_{24}$,
so their contraction is the physical-vector sewing.

\subsection{Six gluons on a mixed-spectrum rectangle}
\label{app:mixed-six-point}

Consider
\begin{equation}
 A_6(1_g,2_g,3_g,4_g,5_g,6_g),\qquad
 I=\{1,2\},\quad b=3,\quad J=(4,5),\quad a=6,
 \label{eq:app-c-six-blocks}
\end{equation}
on the bosonic-string rectangles
\begin{equation}
 h^{(q)}:\qquad s_{14}=s_{15}=2,\qquad
 s_{24}=q,\qquad s_{25}=0,\qquad q=3,2,1,0,-1.
 \label{eq:app-c-six-mixed-family}
\end{equation}
The pairs $14,15$ are nonspectral, so only
the $24$ and $25$ channels can contribute.
For columns $(12),(21)$ and rows $(1,(2)),(2,(1))$,
the insertion matrix is
\[
 \mathsf S_{\sin}
 =\begin{pmatrix}
 \sin\pi(s_{12}+s_{13})&\sin(\pi s_{13})\\
 \sin(\pi s_{23})&\sin\pi(s_{12}+s_{23})
 \end{pmatrix}.
\]
The determinant is
\[
 \det\mathsf S_{\sin}
 =\sin(\pi s_{12})\sin\pi(s_{12}+s_{13}+s_{23}).
\]
On the admissible domain with $\det\mathsf S_{\sin}\ne0$,
inversion and physical sewing give
\begin{equation}
 \begin{aligned}
 &\left.A_6(1_g,2_g,3_g,4_g,5_g,6_g)\right|_{h^{(q)}}
 \\
 &\quad=
 \frac{\pi\alpha'\sin(\pi s_{13})}
 {\sin(\pi s_{12})\sin\pi(s_{12}+s_{13}+s_{23})}
 \\
 &\qquad\times\Bigg[
 \sum_{\mathsf a,\mathsf b}
 A_3(2_g,4_g,-K_{24}^{\mathsf b})\,
 (G_{\rm kin}^{-1})_{\mathsf b\mathsf a}\,
 \left.A_5(1_g,3_g,K_{24}^{\mathsf a},5_g,6_g)
 \right|_{h^{(q)}}
 \\
 &\qquad\qquad
 +(-1)^q
 \sum_{\mathsf a,\mathsf b}
 A_3(2_g,5_g,-K_{25}^{\mathsf b})\,
 (G_{\rm kin}^{-1})_{\mathsf b\mathsf a}\,
 \left.A_5(1_g,3_g,4_g,K_{25}^{\mathsf a},6_g)
 \right|_{h^{(q)}}
 \Bigg].
 \end{aligned}
 \label{eq:app-c-six-factorization}
\end{equation}
The $25$ channel always selects a massless vector;
the $24$ channel follows the four-point spectrum above.
For $q=3,2$, the $24$ fusion space is empty,
so only the $25$ sewing term contributes.
The nonspectral integer $q$ still enters its
transport sign $(-1)^q$.
The surviving five-point factor is a complete
on-shell amplitude.

The gluon bridges
\begin{equation}
 \epsilon_2\cdot k_j=k_2\cdot\epsilon_j
 =\epsilon_2\cdot\epsilon_j=0,
 \qquad j=4,5,
 \label{eq:app-c-six-gluon-bridges}
\end{equation}
remove all selected cubic couplings for
$q=3,2,1,0$, giving
\begin{equation}
 \left.A_6(1_g,2_g,3_g,4_g,5_g,6_g)\right|_{h^{(q)}}=0,
 \qquad q=3,2,1,0.
 \label{eq:app-c-six-hidden-zeros}
\end{equation}
For $q=3,2$, only the $j=5$ bridges are needed.
At $q=-1$, the first-massive $24$ coupling
precludes a nontrivial solution of the pairwise
cubic-zero conditions, as in the four-point case.

\section{KLT kernel zeros and fusion-pole removal}
\label{app:closed-klt}

We specify the Kawai--Lewellen--Tye (KLT) conventions,
prove the rectangular kernel zero, and explain how
the chiral cubic conditions remove compensating poles.

\subsection{KLT conventions and the rectangular kernel zero}
\label{app:klt-bases}

The chiral disk amplitudes use
\begin{equation}
 \alpha'_o=\frac{\alpha'_c}{4},\qquad
 s^c_{ij}=2\alpha'_o k_i\cdot k_j
 =\frac{\alpha'_c}{2}k_i\cdot k_j.
 \label{eq:app-e-slope}
\end{equation}
On the rectangle $h_m^c(\mathbf q)$,
$s^c_{ij}=q_{ij}\in\mathbb Z$ for $i\in I$, $j\in J$;
write $\delta_{ij}=s^c_{ij}-q_{ij}$.
Choose $i_*\in I$ and set
$U=(I\setminus\{i_*\})\cup J$.
The complete KLT word sets are
\begin{equation}
 w_\sigma=(i_*,\sigma(U),b,a),\qquad
 \widetilde w_\gamma=(b,a,\gamma(U),i_*),
 \qquad \sigma,\gamma\in S(U).
 \label{eq:app-e-klt-words}
\end{equation}
Each disk vector has $(n-3)!$ components.
For $\gamma(U)=(u_1,\ldots,u_{|U|})$, the sine
momentum kernel is \cite{Bjerrum-Bohr:2010pnr}
\begin{equation}
 \mathcal K_{\sigma\gamma}
 =\prod_{t=1}^{|U|}\sin\pi\left[
 s^c_{i_*u_t}
 +\sum_{v=t+1}^{|U|}
 \theta_\sigma(u_t,u_v)s^c_{u_tu_v}
 \right],
 \label{eq:app-e-klt-kernel}
\end{equation}
where $\theta_\sigma(u_t,u_v)=1$ if $u_v$ precedes
$u_t$ in $\sigma(U)$, and zero otherwise.
For raw disk periods, our sphere orientation
$d^2z=d\operatorname{Re}z\,d\operatorname{Im}z$
gives \cite{Kawai:1985xq,Bjerrum-Bohr:2010pnr}
\begin{equation}
 \mathcal M_n^{\rm raw}
 =(-1)^{n-3}\sum_{\sigma,\gamma}
 \mathcal A_n^L(w_\sigma)\,
 \mathcal K_{\sigma\gamma}\,
 \mathcal A_n^R(\widetilde w_\gamma).
 \label{eq:app-e-raw-klt}
\end{equation}
Chiral fermions use the fixed reference orders,
with the Koszul sign for collecting left and right
slots included in the sphere convention.
For sums of factorized chiral states, impose the
sufficient cubic conditions term by term.

Choose the last $J$ label in $\gamma(U)$,
say $u_{t_*}=j$.
Every later label lies in $I\setminus\{i_*\}$,
so its sine factor becomes
\begin{equation}
 \sin\pi\left[
 q_{i_*j}
 +\sum_{v=t_*+1}^{|U|}
 \theta_\sigma(j,u_v)q_{u_vj}
 \right]=0.
 \label{eq:app-e-last-j}
\end{equation}
Thus
\begin{equation}
 \left.\mathcal K_{\sigma\gamma}\right|_{h_m^c(\mathbf q)}=0
 \qquad\text{for every }\sigma,\gamma.
 \label{eq:app-e-kernel-zero}
\end{equation}
This extends the rectangular field-theory kernel
argument \cite{Bartsch:2024amu,Li:2024qfp}
to integer sine kernels.

\subsection{Removal of fusion poles}
\label{app:klt-pole-removal}
\label{app:klt-four-point}

The selected chiral fusion mass satisfies
\begin{equation}
 M_{K_{ij}}^2=M_i^2+M_j^2-\frac{q_{ij}}{\alpha'_o},
 \qquad
 \delta_{ij}=\alpha'_o(K_{ij}^2+M_{K_{ij}}^2).
 \label{eq:app-e-chiral-fusion-mass}
\end{equation}
Physical factorization expresses each selected
disk residue as a chiral three-point coupling
sewn to a lower-point amplitude.
Imposing Eq.~\eqref{eq:klt-two-chiral-cubics}
along each fusion divisor therefore removes
these residues in both chiral sectors.
Nonspectral pairs require no cubic condition.

Choose a rectangular point away from all other
physical poles, as assumed in the main text.
In a sufficiently small neighborhood, the only
possible poles are the selected fusion poles
$\delta_{ij}=0$, each of at most simple order. 
We can therefore write
\[
 \mathcal A_X
 =\frac{F_X}
 {\displaystyle\prod_{i\in I}\prod_{j\in J}\delta_{ij}},
 \qquad X=L,R,
\]
where $F_X$ is regular near the chosen point.
The persistent cubic-zero conditions in
Eq.~\eqref{eq:klt-two-chiral-cubics}
remove every selected spectral pole.
Absent channels and nonspectral pairs are already
regular, so every denominator factor cancels.
Since the $\delta_{ij}$ are independent local
coordinates, these cancellations hold jointly:
both disk vectors remain regular as all
$\delta_{ij}\to0$.
The entrywise kernel zero in
Eq.~\eqref{eq:app-e-kernel-zero} then gives
\[
 \left.\mathcal M_n\right|_{h_m^c(\mathbf q)}=0.
\]
The regular disk values need not vanish, and no
inverse of the open insertion matrix is needed.
The cubic conditions \eqref{eq:klt-two-chiral-cubics} must hold with the other
parameters free: pointwise zeros can leave residues
proportional to another parameter $\tau$, producing
a term proportional to $\tau^2/\delta$ already
at four points.

To illustrate the mechanism, consider four massless
closed bosonic-string states at $q_{13}=0$.
Set $2\alpha'_o=1$, so $\alpha'_c=2$, and choose
a complex on-shell family with nonzero $K_{13}$ and
\[
 \delta=s^c_{13},\qquad
 s^c_{23}=-s^c_{12}-\delta.
\]
Let $u,v$ be fixed orthogonal unit vectors transverse
to all momenta throughout this family.
In both chiralities choose
\[
 (\epsilon_1,\epsilon_2,\epsilon_3,\epsilon_4)_X
 =(u,u,v,v),\qquad X=L,R.
\]
The factorized closed-string polarizations
$\epsilon_{r,L}\otimes\epsilon_{r,R}$ are physical
and nonzero. They contain trace components and
are not pure-graviton polarizations.

These bridges annihilate the complete $13$ vector
cubic functionals in both chiral sectors along
$\delta=0$.
Physical factorization therefore removes the
singular parts of both disk amplitudes.
Their complete expressions are
\[
 \mathcal A_L(1324)
 =B(1+\delta,1-s^c_{12}-\delta),
 \qquad
 \mathcal A_R(2431)
 =B(1+\delta,s^c_{12}-1),
\]
with nonzero regular values
\[
 \left.\mathcal A_L\right|_{\delta=0}
 =\frac1{1-s^c_{12}},
 \qquad
 \left.\mathcal A_R\right|_{\delta=0}
 =-\frac1{1-s^c_{12}}.
\]
The KLT relation then gives
\begin{equation}
 \mathcal M_4^{\rm raw}
 =-\sin(\pi\delta)\,
 \mathcal A_L(1324)\mathcal A_R(2431)
 =O(\delta).
 \label{eq:app-e-four-klt}
\end{equation}
The cubic-zero conditions remove the disk poles,
while the kernel supplies the sphere zero.

\section{Worldsheet OPEs and relative-scale decomposition}
\label{app:ope-scale}

We give local OPE realizations of cross regularity,
derive the relative-scale measures, and illustrate
the scale picture with four-point Euler periods.

\subsection{Local OPE inputs}
\label{app:local-ope-inputs}

For massless bosonic gluons, the bridges in
Eq.~\eqref{eq:app-c-zero-gluon-bridges}
remove all oscillator cross contractions.
Including the plane-wave factor
$(z_i-z_j)^{q_{ij}}$, the remaining product is
regular for $q_{ij}\ge0$.

For canonical massless Ramond--Neveu--Schwarz vertices
\cite{Friedan:1985ge,Schlotterer:2010kk} at $q_{ij}=0$,
the corresponding bridges remove the principal parts
of $g^{(0)}g^{(0)}$, $g^{(-1)}g^{(0)}$,
$g^{(0)}f^{(-1/2)}$ and
$f^{(-1/2)}f^{(-1/2)}$.
In particular, the spin-field OPE
\cite{Haertl:2010qlb} gives
\begin{equation}
 V_{f_i}^{(-1/2)}(w+t)V_{f_j}^{(-1/2)}(w)
 \sim\frac{\bar\chi_i\Gamma^\mu\chi_j}{t}\,
 V_{g,\mu}^{(-1)}(K_{ij};w),
 \qquad t=z_i-z_j,
 \label{eq:app-f-fermion-ope}
\end{equation}
where only the singular term is displayed and a
nonzero normalization is suppressed.
The vector-current bridge therefore removes this term;
the mixed bridges work similarly
\cite{Zhang:2026wye}.

These OPEs are used in a common prescription of total
picture $-2$.
With no Ramond states, place the two picture-$-1$
NS vertices on $a,b$.
With two Ramond states at picture $-1/2$, use one
picture-$-1$ NS vertex, on a distinguished NS leg
when available. If both distinguished legs are
Ramond, all active cross pairs are NS--NS.
With four Ramond states, keep all four at picture
$-1/2$ and every NS vertex at picture zero.
All unspecified NS vertices have picture zero.
For higher Ramond multiplicity, include the required
picture-changing contributions in the complete OPE
check \cite{Berkovits:1997mc}.
On the sphere, apply the prescription separately
in each chirality.

\paragraph{Leading-Regge tensors.}
\label{app:massive-tensor-ope}

The tensor rule in
Eq.~\eqref{eq:app-c-zero-regge-rule}
follows from the standard vertex
\cite{Sagnotti:2010at}
\begin{equation}
 V_i=
 :E_{i,\mu_1\cdots\mu_{L_i}}
 \partial X^{\mu_1}\cdots
 \partial X^{\mu_{L_i}}e^{ik_i\cdot X}:.
\end{equation}
Under the momentum bridges
$E_i\cdot k_j=E_j\cdot k_i=0$,
Wick contraction gives
\begin{equation}
 V_i(w+t)V_j(w)
 =
 \sum_{r=0}^{\min(L_i,L_j)}
 (-2\alpha')^r r!
 \binom{L_i}{r}\binom{L_j}{r}
 t^{q_{ij}-2r}\mathcal N_r(t).
 \label{eq:app-f-regge-ope}
\end{equation}
For $r\ge1$, $\mathcal N_r(t)$ is linear in the
partial contraction $C_r$ and has a regular Taylor
expansion; $r=0$ contains the uncontracted tensor
product.

For $q_{ij}\ge0$, negative powers require
$r\ge r_0=\lfloor q_{ij}/2\rfloor+1$.
Setting $C_{r_0}=0$ also removes its further
contractions, so every singular term vanishes.
If $r_0>\min(L_i,L_j)$, the momentum bridges
alone suffice.
For $L_i=L_j=2$, the three coefficients are
$1,-8\alpha',8\alpha'^2$.
At $q_{ij}=2,3$, only the double contraction
can be singular, giving $E_i:E_j=0$.

\subsection{Relative-scale coordinates and measures}
\label{app:relative-scale}

Use the finite rational representation and
cross-regularity conditions specified in the main text.
After cross-denominator cancellation, the remaining
denominators scale homogeneously.
Any picture-changing contributions must be included
in the same homogeneous prescription
\cite{Sen:2015hia}.

In the gauge $z_b=0$, $z_a=\infty$, $z_m=-1$,
the right-block substitution is
\begin{equation}
 (z_{j_1},\ldots,z_{j_{|J|}})
 =(\lambda x_1,\ldots,\lambda x_{|J|-1},\lambda),
 \qquad x_{|J|}=1.
 \label{eq:app-d-scale-coordinates}
\end{equation}
Its Jacobian is $\lambda^{|J|-1}$.
The Faddeev--Popov factor is constant after the
normalized limit at infinity, giving the ordered
coordinate density
\begin{equation}
 d\mu_{\rm ord}
 =\lambda^{|J|-1}d\lambda
 \prod_{t=1}^{|J|-1}dx_t
 \prod_{i=1}^{m-1}dy_i.
 \label{eq:app-d-ordered-density}
\end{equation}

The Koba--Nielsen (KN) factors internal to
$J\cup\{b\}$ contribute the degree
\begin{equation}
 \beta_J
 =\sum_{t=1}^{|J|}s_{bj_t}
 +\sum_{1\le t<u\le |J|}s_{j_tj_u}.
 \label{eq:app-d-kn-degree}
\end{equation}
The polynomial cross factors
$(\lambda x_t+y_i)$ and the remaining homogeneous
factors supply finitely many integer shifts $r_\ell$.
Together with the measure, this gives
\begin{equation}
 \Omega_{\rm rect}^{\rm open}
 =\frac{d\lambda}{\lambda}
 \sum_\ell\lambda^{\Delta_\ell}\widehat\Omega_\ell,
 \qquad
 \Delta_\ell=|J|+\beta_J+r_\ell.
 \label{eq:app-d-scale-decomposition}
\end{equation}
The forms $\widehat\Omega_\ell$ depend only on the
shape variables and external-state data.
Related polynomial structures occur in curve-integral
descriptions of string zeros
\cite{Arkani-Hamed:2023swr,Chang:2025cqe}.

On the sphere, the same substitution uses complex
$x,y$ and $\lambda\in\mathbb C^\times$.
The coordinate density becomes
\begin{equation}
 d\mu_{\rm sphere}
 =|\lambda|^{2(|J|-1)}d^2\lambda
 \prod_{t=1}^{|J|-1}d^2x_t
 \prod_{i=1}^{m-1}d^2y_i.
 \label{eq:app-f-sphere-jacobian}
\end{equation}
Let $\beta_J^c$ denote Eq.~\eqref{eq:app-d-kn-degree}
with $s$ replaced by $s^c$.
The two chiral factors then give
\begin{equation}
 \Omega_{\rm rect}^{\rm closed}
 =\frac{d^2\lambda}{|\lambda|^2}
 \sum_\ell
 \lambda^{h_\ell}\bar\lambda^{\widetilde h_\ell}
 \widehat\Omega_\ell,
 \qquad
 h_\ell=|J|+\beta_J^c+r_\ell,\quad
 \widetilde h_\ell=|J|+\beta_J^c+\widetilde r_\ell.
 \label{eq:app-f-closed-scale-decomposition}
\end{equation}
At fixed shape, cross regularity removes the
finite-scale collisions $\lambda=-y_i/x_t$,
leaving the scale endpoints and shape boundaries.

\subsection{Scaleless sectors and four-point periods}
\label{app:four-point-scale-check}

For an open-string sector, consider
\begin{equation}
 \mathcal S(\Delta)
 =\int_0^\infty\frac{d\lambda}{\lambda}\lambda^\Delta.
 \label{eq:app-d-scaleless-period}
\end{equation}
Splitting at $\lambda_*>0$, the two endpoint
integrals continue to
$\lambda_*^\Delta/\Delta$ and
$-\lambda_*^\Delta/\Delta$.
Their sum vanishes, including at $\Delta=0$
after combining the continued expressions.

For the sphere, $\lambda=Re^{i\theta}$ gives
\begin{equation}
 \frac{d^2\lambda}{|\lambda|^2}
 \lambda^h\bar\lambda^{\widetilde h}
 =
 \frac{dR}{R}\,d\theta\,
 R^{h+\widetilde h}
 e^{i(h-\widetilde h)\theta}.
 \label{eq:app-f-angular-radial-period}
\end{equation}
Single-valuedness requires
$h-\widetilde h\in\mathbb Z$.
Angular integration removes nonzero angular weights;
the surviving sectors carry
$2\pi\mathcal S(2h)$.

Simple four-point Euler periods illustrate how this
scale prescription agrees with continuation of a
complete integral. These checks use raw disk periods at
$2\alpha'=1$ and chiral slope $2\alpha'_o=1$
for the sphere.
For the bosonic-gluon component $(u,u,v,v)$ in
Appendix~\ref{app:klt-four-point}, the complete raw
disk period $\mathcal I_{\rm open}=\mathcal A_4(1234)$ is
\begin{equation}
 \begin{aligned}
 \mathcal I_{\rm open}
 &=
 \int_0^\infty d\lambda\,
 \lambda^{-s_{12}-s_{13}}(1+\lambda)^{s_{13}}
=
 \frac{\Gamma(s_{12}-1)
 \Gamma(1-s_{12}-s_{13})}{\Gamma(-s_{13})}.
 \end{aligned}
 \label{eq:app-d-four-gluon-period}
\end{equation}
Setting $s_{13}=N\in\mathbb N_0$ makes
$(1+\lambda)^N$ a finite polynomial.
Each monomial gives a scaleless sector, while
the complete continued period vanishes through
$1/\Gamma(-N)=0$ for generic noninteger $s_{12}$.

For the physical states and invariant family of
Appendix~\ref{app:klt-four-point},
the gauge $(z_1,z_2,z_4)=(-1,0,\infty)$,
with $\lambda=z_3$ and $\delta=s^c_{13}$,
gives the complete raw sphere period
$\mathcal I_{\rm closed}=\mathcal M_4^{\rm raw}$:
\begin{equation}
 \begin{aligned}
 \mathcal I_{\rm closed}
 &=
 \int_{\mathbb C}d^2\lambda\,
 |\lambda|^{-2s^c_{12}-2\delta}|1+\lambda|^{2\delta}
=
 \pi
 \frac{
 \Gamma(s^c_{12}-1)
 \Gamma(1-s^c_{12}-\delta)\Gamma(1+\delta)}
 {\Gamma(2-s^c_{12})
 \Gamma(s^c_{12}+\delta)\Gamma(-\delta)}.
 \end{aligned}
 \label{eq:app-f-four-closed-period}
\end{equation}
This is the complex Euler continuation
\cite{Brown:2019wna}.
At $\delta=0$, its integrand becomes
\begin{equation}
 d^2\lambda\,|\lambda|^{-2s^c_{12}}
 =\frac{dR}{R}\,d\theta\,R^{2(1-s^c_{12})},
 \label{eq:app-f-four-closed-scaleless}
\end{equation}
with zero angular weight and one scaleless radial
sector.
The complete continued expression likewise vanishes
through $1/\Gamma(-\delta)$ for generic noninteger
$s^c_{12}$.

At higher multiplicity, matching this sectorwise
prescription to the complete periods requires control
of overlapping scale and shape degenerations
\cite{Mizera:2017cqs,Brown:2019wna}.
\end{document}